# CHIME Discovery of a Binary Pulsar with a Massive Nondegenerate Companion

Bridget C. Andersen[1,2], Emmanuel Fonseca[3,4], J. W. McKee[5,6,7], B. W. Meyers[8,9], Jing Luo[5], C. M. Tan[1,2], I. H. Stairs[8], Victoria M. Kaspi[1,2], M. H. van Kerkwijk[10], Mohit Bhardwaj[1,2], P. J. Boyle[1,2], Kathryn Crowter[8], Paul B. Demorest[11], Fengqiu A. Dong[8], Deborah C. Good[12,13], Jane F. Kaczmarek[14], Calvin Leung[15,16], Kiyoshi W. Masui[15,16], Arun Naidu[1,2], Cherry Ng[17], Chitrang Patel[1,2], Aaron B. Pearlman[1,2], Ziggy Pleunis[17], Masoud Rafiei-Ravandi[1,2], Mubdi Rahman[18], Scott M. Ransom[19], Kendrick M. Smith[20], and Shriharsh P. Tendulkar[21,22]

[1] Department of Physics, McGill University, 3600 rue University, Montréal, QC H3A 2T8, Canada; bridget.andersen@mail.mcgill.ca
[2] McGill Space Institute, McGill University, 3550 rue University, Montréal, QC H3A 2A7, Canada
[3] Department of Physics and Astronomy, West Virginia University, PO Box 6315, Morgantown, WV 26506, USA
[4] Center for Gravitational Waves and Cosmology, West Virginia University, Chestnut Ridge Research Building, Morgantown, WV 26505, USA
[5] Canadian Institute for Theoretical Astrophysics, 60 St. George Street, Toronto, ON M5S 3H8, Canada
[6] E. A. Milne Centre for Astrophysics, University of Hull, Cottingham Road, Kingston-upon-Hull, HU6 7RX, UK
[7] Centre of Excellence for Data Science, Artificial Intelligence and Modelling (DAIM), University of Hull, Cottingham Road, Kingston-upon-Hull, HU6 7RX, UK
[8] Department of Physics and Astronomy, University of British Columbia, 6224 Agricultural Road, Vancouver, BC V6T 1Z1 Canada
[9] International Centre for Radio Astronomy Research (ICRAR), Curtin University, 1 Turner Avenue, Technology Park, Bentley, 6102, WA, Australia
[10] David A. Dunlap Department of Astronomy & Astrophysics, University of Toronto, 50 St. George Street, Toronto, ON M5S 3H4, Canada
[11] National Radio Astronomy Observatory, P.O. Box O, Socorro, NM 87801, USA
[12] Department of Physics, University of Connecticut, 196A Auditorium Road Unit 3046, Storrs, CT, USA
[13] Center for Computational Astrophysics, Flatiron Institute, 162 5th Avenue, New York, NY 10010, USA
[14] Dominion Radio Astrophysical Observatory, Herzberg Research Centre for Astronomy and Astrophysics, National Research Council Canada, P.O. Box 248, Penticton, BC V2A 6J9, Canada
[15] MIT Kavli Institute for Astrophysics and Space Research, Massachusetts Institute of Technology, 77 Massachusetts Avenue, Cambridge, MA 02139, USA
[16] Department of Physics, Massachusetts Institute of Technology, 77 Massachusetts Avenue, Cambridge, MA 02139, USA
[17] Dunlap Institute for Astronomy & Astrophysics, University of Toronto, 50 St. George Street, Toronto, ON M5S 3H4, Canada
[18] Sidrat Research, P.O. Box 73527 RPO Wychwood, Toronto, ON M6C 4A7, Canada
[19] National Radio Astronomy Observatory, 520 Edgemont Road, Charlottesville, VA 22903, USA
[20] Perimeter Institute for Theoretical Physics, 31 Caroline Street N, Waterloo, ON N25 2YL, Canada
[21] Department of Astronomy and Astrophysics, Tata Institute of Fundamental Research, Mumbai, 400005, India
[22] National Centre for Radio Astrophysics, Post Bag 3, Ganeshkhind, Pune, 411007, India
Received 2022 September 13; revised 2022 November 16; accepted 2022 November 19; published 2023 January 24

## Abstract

Of the more than 3000 radio pulsars currently known, only ∼300 are in binary systems, and only five of these consist of young pulsars with massive nondegenerate companions. We present the discovery and initial timing, accomplished using the Canadian Hydrogen Intensity Mapping Experiment (CHIME) telescope, of the sixth such binary pulsar, PSR J2108+4516, a 0.577 s radio pulsar in a 269 day orbit of eccentricity 0.09 with a companion of minimum mass 11 $M_\odot$. Notably, the pulsar undergoes periods of substantial eclipse, disappearing from the CHIME 400–800 MHz observing band for a large fraction of its orbit, and displays significant dispersion measure and scattering variations throughout its orbit, pointing to the possibility of a circumstellar disk or very dense stellar wind associated with the companion star. Subarcsecond resolution imaging with the Karl G. Jansky Very Large Array unambiguously demonstrates that the companion is a bright, $V \simeq 11$ OBe star, EM* UHA 138, located at a distance of 3.26(14) kpc. Archival optical observations of EM* UHA 138 approximately suggest a companion mass ranging from 17.5 $M_\odot < M_c < 23\ M_\odot$, in turn constraining the orbital inclination angle to $50°.3 \lesssim i \lesssim 58°.3$. With further multiwavelength follow-up, PSR J2108+4516 promises to serve as another rare laboratory for the exploration of companion winds, circumstellar disks, and short-term evolution through extended-body orbital dynamics.

*Unified Astronomy Thesaurus concepts:* Radio pulsars (1353); Binary pulsars (153); Compact binary stars (283); Pulsars (1306); Neutron stars (1108); Be stars (142); Circumstellar disks (235)

## 1. Introduction

Of the more than 3000 radio pulsars currently cataloged (Hobbs et al. 2004),[23] only five are presently known to be in binaries with massive, nondegenerate companions (Johnston et al. 1992; Kaspi et al. 1994; Stairs et al. 2001; Lorimer et al. 2006; Lyne et al. 2015). Pulsar–massive-star binaries are thought to represent an intermediate stage in the binary evolution of two high-mass stars, where the initially more massive star has undergone a core-collapse supernova to form a neutron star (e.g., Bhattacharya & van den Heuvel 1991; Phinney & Kulkarni 1994). If the binary remains bound following this explosion, and if the resulting neutron star is observable as a radio pulsar, such a pulsar/massive-star binary is born. These systems can be phenomenologically rich, as the pulsar and nondegenerate stellar winds can interact, and the pulsar radio emission can be affected by enhanced orbital-phase dependent dispersion, scattering, and eclipsing. As such, these systems are rare laboratories for the exploration of massive star and pulsar winds (e.g., Kaspi et al. 1996), circumstellar disks (e.g., Melatos et al. 1995), and extended-body orbital dynamics (e.g., Lai et al. 1995).

---

[23] See the ATNF pulsar catalog at https://www.atnf.csiro.au/research/pulsar/psrcat/.

PSR B1259−63 was the first pulsar/massive-star binary system discovered, and consists of a young, 47 ms radio pulsar in a 3.4 yr highly eccentric orbit with a $\sim 8\,M_\odot$ Be star (Johnston et al. 1992). The pulsar radio emission is highly scattered and undergoes an eclipse at periastron passages, and the interaction between the winds of the two components results in high-energy emission from X-ray (e.g., Kaspi et al. 1995) up to TeV energies (Abdo et al. 2011; Aharonian et al. 2005). This source has been useful for studying pulsar/ massive-stellar wind interactions and has helped elucidate the nature of other gamma-ray binaries in which no radio pulsations have yet been observed (e.g., Chernyakova et al. 2006). The second discovered pulsar/massive-star system was PSR J0045−7319, a 0.9 s pulsar in a 51 day highly eccentric orbit with a B star, located in the Small Magellanic Cloud (Kaspi et al. 1994). In this case, the stellar wind is weak (Kaspi et al. 1996), and no scattering or eclipsing is observed in spite of the close proximity of the two components at periastron (just four companion radii, or about $25\,R_\odot$). On the other hand, strong dynamical spin–orbit coupling is observed, a result of the spin-induced quadrupolar moment of the B star, and a misalignment between the B star's spin and the orbital plane (Kaspi et al. 1996). Another known radio pulsar/massive-star binary is PSR J1740−3052 (Stairs et al. 2001; Bassa et al. 2011; Madsen et al. 2012), which consists of a 0.57 s pulsar in a highly eccentric 231 day binary orbit with a B star. This source shows dispersion measure (DM) and scattering variations presumably due to its companion's wind, as well as interesting dynamical effects like spin–orbit coupling. J2032 +4127 is a 0.14 s pulsar in a 45–50 yr highly eccentric orbit with a massive Be star (Lyne et al. 2015; Ho et al. 2017) that produces broadband emission from radio to TeV gamma rays (Chernyakova et al. 2020) due to the interaction of the pulsar wind with the mass outflow from the companion, likely in an inclined orbit–wind geometry. Finally PSR J1638−4725 is a 0.76 s pulsar in a 5.3 yr (1940 day) binary with an as-yet unknown companion (Lorimer et al. 2006); little else has been published about this system.[24]

Such binaries are rare, but modern pulsar searches have greater sensitivity than previous efforts and therefore offer new discovery opportunities. The Canadian Hydrogen Intensity Mapping Experiment (CHIME) is a radio telescope that possesses a very wide field of view, large collecting area and high sensitivity across the 400–800 MHz range. One of the commensal backends on the CHIME telescope is the CHIME/ FRB instrument (CHIME/FRB Collaboration et al. 2018), which autonomously detects extragalactic fast radio bursts (FRBs) in real time. However, Galactic sources—single pulses from pulsars and "rotating radio transients"—can also be detected by the CHIME/FRB backend, making CHIME/FRB a Galactic pulsar discovery instrument as well (e.g., Good et al. 2021). In particular, CHIME's large field of view and daily observing cadence makes it uniquely primed for detecting eclipsing or otherwise disappearing systems like some pulsar/ massive-star binaries. Pulsars discovered by CHIME/FRB can then be studied in detail using the CHIME/Pulsar backend (CHIME/Pulsar Collaboration 2021), a separate computing system that acquires data appropriate for pulsar timing and high-resolution search experiments.

In this paper, we present the CHIME/FRB discovery and CHIME/Pulsar study of PSR J2108+4516, a 0.577 s radio pulsar in a 269 day orbit with a massive nondegenerate star. The initial discovery of this source was first reported in Good et al. (2021).[25] Section 2 describes the CHIME/FRB and CHIME/Pulsar observations used in our discovery and study of PSR J2108+4516. Section 3 describes our determination of the orbit and timing analysis methods. Section 4 describes our observations with the Jansky Very Large Array (VLA), which localized PSR J2108+4516 on the sky, and revealed a coincidence with a previously cataloged massive nondegenerate companion star. Section 5 discusses the results, including large-scale DM and scattering variations likely due to the massive star's wind. Finally, Section 6 presents our conclusions.

## 2. CHIME Observations

We discovered and initially monitored PSR J2108+4516 with the CHIME telescope, using the CHIME/FRB and CHIME/Pulsar backends to acquire various types of data. The CHIME acquisition and processing methods used in this work were similar to those presented by Good et al. (2021), and we provide a summary below.

### 2.1. CHIME/FRB

CHIME/FRB continuously receives 1024 total-intensity data streams from the CHIME correlator with time and frequency resolutions of 0.983 ms and 24.4 kHz, respectively. Each stream corresponds to a static, tied-array beam that points to a fixed altitude and azimuth. These beams are positioned to spatially tile the CHIME primary beam extending $\sim 2°$ east–west and $\sim 120°$ north–south, allowing CHIME/FRB to scan the entire sky north of decl. $-11°$ for roughly 15 minutes daily. The CHIME/FRB search pipeline runs in real time on the time series data from each of these formed beams. In short, the pipeline cleans the drift-scan time series of radio frequency interference (RFI), searches the cleaned time series for dispersed pulses, categorizes candidate pulses based on derived metrics and metadata, and records one or more data products based on configurable signal-to-noise ratio (S/N) thresholds for data management. Coarse metadata for each candidate pulse, such as the detection S/N, position, time of arrival (TOA), and DM, are stored in the CHIME/FRB database. More details about the pipeline can be found in CHIME/FRB Collaboration et al. (2018).

On 2018 October 11 (MJD 58402), the CHIME/FRB backend first detected a transit of 168 individual pulsations from PSR J2108+4516 above the CHIME/FRB detection S/N threshold of 8. These initial detections were classified by the pipeline as arising from an unknown Galactic source, as its tentative position and coarse DM were consistent with placement within the Milky Way when compared to available models of free Galactic electrons (Cordes & Lazio 2002; Yao et al. 2017). In late 2018, the real-time CHIME/FRB pipeline was configured to dump short spans of intensity data for unknown Galactic pulses with detection S/N greater than 10. For example, on 2018 October 13 (MJD 58404), a pulse from PSR J2108+4516 reached this threshold, and the CHIME/FRB system recorded ∼13 s of intensity data in which multiple pulses were visible. A section of this data is shown in Figure 1. We leveraged these data to obtain initial estimates of the spin period and DM, yielding ∼0.577 s and ∼82.5 pc cm$^{-3}$, respectively.

---

[24] Recently, Parent et al. (2022) also identified an interesting pulsar, J1954 +2529, which might have a low-mass nondegenerate companion. But the true nature of this candidate has yet to be confirmed.

[25] See the CHIME/FRB Galactic sources webpage: https://www.chime-frb. ca/galactic.

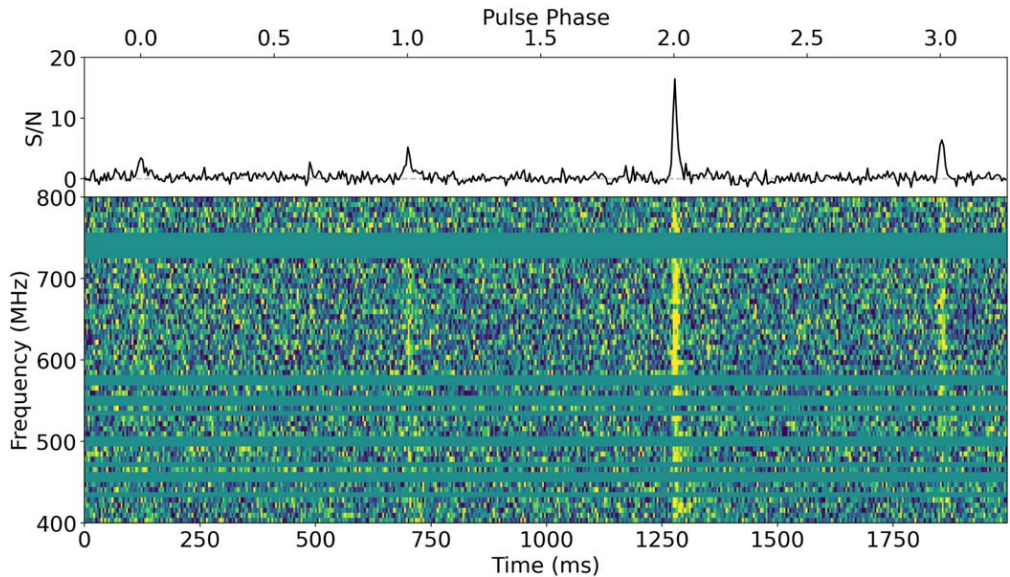

**Figure 1.** A 2 s section of CHIME/FRB intensity data from an early transit of PSR J2108+4516 on 2018 October 13 (MJD 58404). The data has been dedispersed to 83.5 pc cm$^{-3}$ and downsampled to 4 ms time resolution and 6.25 MHz frequency resolution (64 subbands) for plotting clarity. The highest-S/N pulse triggered the real-time intensity dump. The horizontal gray dashed line in the top panel denotes 0 S/N. The blank horizontal sections in the bottom panel indicate frequency channels that have been flagged due to RFI.

Since the initial detection on MJD 58402, CHIME/FRB has continued daily monitoring of single pulses from PSR J2108 +4516 up to the present day. This monitoring has revealed extended periods of nondetection, where the pulsations, if present, fall below the S/N = 8 detection threshold. Notably on 2018 October 31 (MJD 58422), just 20 days after the initial detection, PSR J2108+4516 disappeared and was not detected again until 2019 February 10 (MJD 58524, 102 days later).[26] The subsequent pattern of CHIME/FRB detections, including several more extended periods of nondetection, is summarized in the top panel of Figure 2. A full discussion of the sensitivity and selection function of the CHIME/FRB pipeline is out of the scope of this paper (for more general information, see Good et al. 2021; and CHIME/FRB Collaboration et al. 2021).

### 2.2. CHIME/Pulsar

The CHIME/Pulsar backend (CHIME/Pulsar Collaboration 2021) is a separate acquisition system for CHIME that receives 10 streams of beamformed, complex channelized voltages with time and frequency resolutions of 2.56 μs and 0.390625 MHz, respectively. Each beamformed time series is computed by the CHIME correlator using time-dependent phases to digitally emulate the tracking of specified radio sources as they traverse the CHIME primary beam. Upon the reception of beamformed baseband, the CHIME/Pulsar backend performs coherent dedispersion (Hankins & Rickett 1975) at a specified DM for each beam and enacts one of two possible downsampling algorithms: (1) a reduction in time resolution suitable for searching experiments; (2) a phase-coherent "folding" of consecutive pulses, based on an existing timing ephemeris, into a set of integrated profiles across the transit that are suitable for high-precision timing studies. We refer to the former mode of observing as a *filterbank* observation, and the latter mode as a *fold-mode* observation.

We began observing PSR J2108+4516 with the CHIME/ Pulsar backend in filterbank mode starting on 2018 October 20 (MJD 58411). A DM of 82.5 pc cm$^{-3}$ was used for coherent dedispersion. For each observation, one of the ten beams was assigned to track the trajectory of PSR J2108+4516 for ∼20 minutes centered around the meridian transit time. This duration encompasses the amount of time it takes a source to fully transit the primary beam at the decl. of PSR J2108+4516. Early observations were scheduled sporadically as the CHIME/Pulsar experiment was still being commissioned. After 2019 January 15 (MJD 58498), we prioritized observations of PSR J2108+4516 to occur at a near-daily cadence (for more on the CHIME/Pulsar scheduling algorithm, see CHIME/Pulsar Collaboration 2021).

Once initial estimates for the position, DM, and spin period were established from CHIME/FRB data, we switched to using the CHIME/Pulsar backend in fold-mode on 2018 November 19 (MJD 58441), generating full-Stokes pulse profiles folded into 10 s subintegrations, again assuming a DM of 82.5 pc cm$^{-3}$. During an interlude lasting from 2019 June 11 (MJD 58645) to 2019 October 22 (MJD 58778), when the source entered an extended period of nondetection for a second time, we switched back to filterbank data to create more flexibility for pulse searches. After this interlude, we returned to fold-mode observations, which continue up until the present day. Detection and nondetection eras with CHIME/Pulsar are shown in the bottom panel of Figure 2. Although polarimetry is available in the filterbank data, we defer the polarization analysis to a future work.

---

[26] The timing of this first period of nondetection earned PSR J2108+4516 the nickname *Snowbird*, in reference to travelers that *disappear* during cold northern winters to live in warmer southern climes.

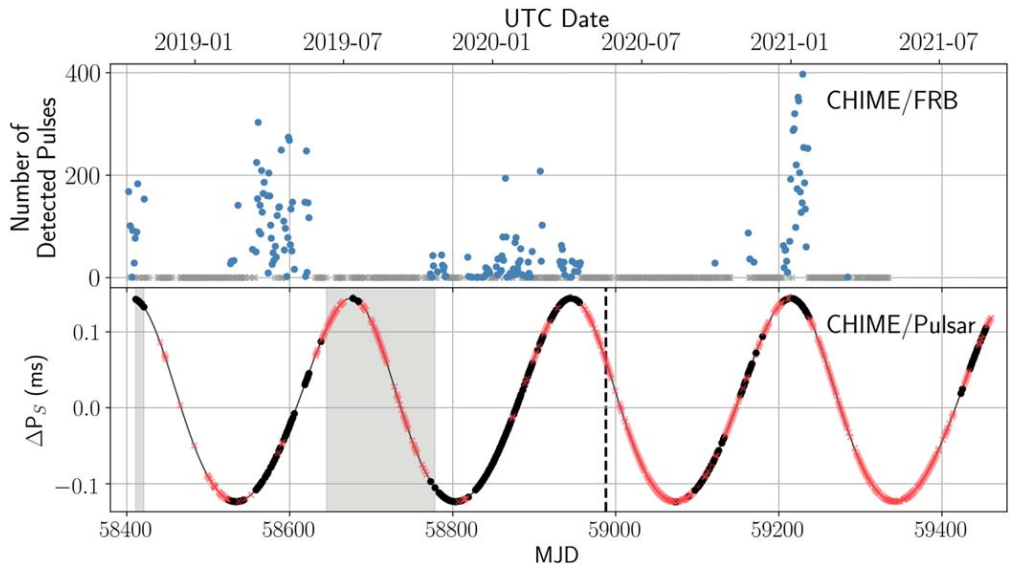

**Figure 2.** A summary of the PSR J2108+4516 data sets obtained from the CHIME/FRB and CHIME/Pulsar backends. Top: the blue circles show the number of CHIME/FRB single pulses that were detected per UTC day up to MJD 59334. Note that this figure only contains detections on days with nonzero exposure when the system was functioning nominally. If there were no detections on a day with nonzero exposure, then a gray x is marked at the zero line of the number of detected pulses. The UTC year and month are labeled on the top axis. Bottom: change in the pulsar spin period as a function of time with CHIME/Pulsar detections and nondetections displayed as black dots and red x's, respectively. The pulsar spin period modulates over time due to the binary motion of PSR J2108+4516. The gray line denotes the best-fit timing estimate of the orbital motion. The grayed-out intervals indicate periods of time when CHIME/Pulsar was configured to take filterbank data. The rest of the observations were taken in fold-mode. The vertical black dashed line indicates MJD 58988, the date of our VLA S-band observation (see Section 4).

## 3. Analysis and Results

We acquired ∼2.8 yr of near-daily CHIME/Pulsar observations of PSR J2108+4516 extending from 2018 October 20 (MJD 58411) to 2021 September 3 (MJD 59460). This data set is the subject of analyses described below and includes 756 CHIME/Pulsar observations: 85 in filterbank and 671 in fold-mode.

An initial examination of these observations promptly revealed PSR J2108+4516 as a physically distinct source with a number of properties:

1. *Binary orbit.* The profile drifts over pulse phase were apparent within the first few CHIME/Pulsar fold-mode observations, indicating that the pulsar was experiencing significant acceleration from orbiting with a binary companion. A further timing analysis (Section 3.4) revealed an orbital period of 269 days.
2. *Significant dispersion and scattering variations.* PSR J2108+4516's pulse profile exhibits extreme simultaneous dispersion and scattering variations on intra-day timescales that are unlike behavior seen in most of the pulsar population. For example, Figure 3 shows the pulse profile variation over a 9 day period in which the DM varied by ∼2 pc cm$^{-3}$, and the scattering time at 1 GHz varied by ∼6.5 ms.
3. *Periods of nondetection and nulling.* The extended periods of nondetection first observed in CHIME/FRB real-time pipeline triggers (Section 2.1) were also apparent in our CHIME/Pulsar observations. Notably, PSR J2108+4516 disappears for 88–186 days at a time at roughly the same orbital phase. Additionally, we observed instances of nulling, where pulses appear to switch on and off within a single observation (as shown in Figure 4).

In the following sections, we further explore and quantify the properties of PSR J2108+4516 based on our analysis of the CHIME/Pulsar data set. Section 3.1 reviews the initial preprocessing steps that we completed on the CHIME/Pulsar data to prepare them for further analysis. Section 3.2 presents the pattern of detections and nondetections and nulling in more detail, and identifies periods of nondetection representing eclipses. Section 3.3 describes how we derive DM measurements, scattering measurements, and TOAs using the Pulse-Portraiture[27] software package (Pennucci et al. 2014; Pennucci 2019). Finally, Section 3.4 describes how we used those PulsePortraiture measurements to obtain a timing solution for PSR J2108+4516.

### 3.1. Folding, RFI Excision, and Downsampling

We used commonly employed preprocessing utilities on each CHIME/Pulsar observation in order to generate cleaned data products for the eventual profile fitting and TOA extraction described below. We summarize these processing steps here, which include data folding, RFI excision, and downsampling.

1. *Folding.* We processed all filterbank data into 10 s subintegrations using the latest timing ephemeris and the Digital Signal Processing Software for Pulsar Astronomy (dspsr; van Straten & Bailes 2011).[28] This folding step was initially completed using the approximate position, spin period, and DM obtained from CHIME/FRB observations, and was then repeated with each improved iteration of the timing solution (which we describe in Section 3.4). We also updated existing CHIME/Pulsar fold-mode data to base the pulse calculations on the latest

---
[27] https://github.com/pennucci/PulsePortraiture
[28] http://dspsr.sourceforge.net/

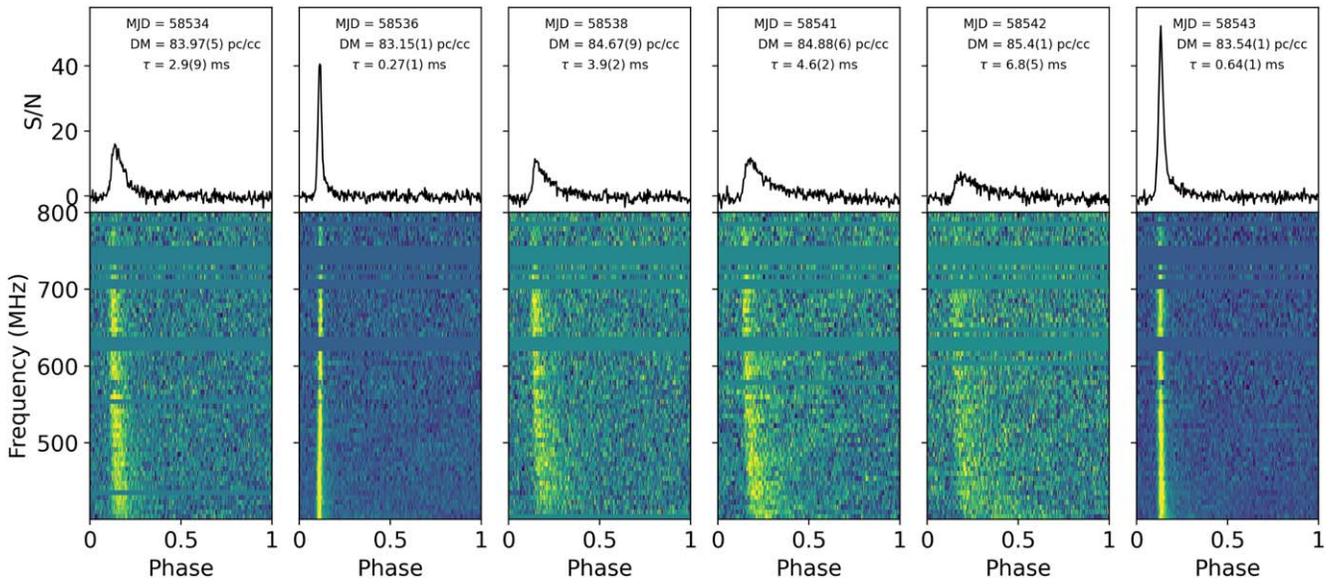

**Figure 3.** Rapid variation in the folded pulse profile of PSR J2108+4516 over a 9 day period. All profiles are dedispersed to 83.54 pc cm$^{-3}$ (the measured DM for MJD 58543) and integrated over the transit time. Best-fit estimates of DM and scattering timescale as determined by `PulsePortraiture` are listed in each profile panel with their 1$\sigma$ uncertainties in parentheses. The scattering timescales are referenced to 1 GHz.

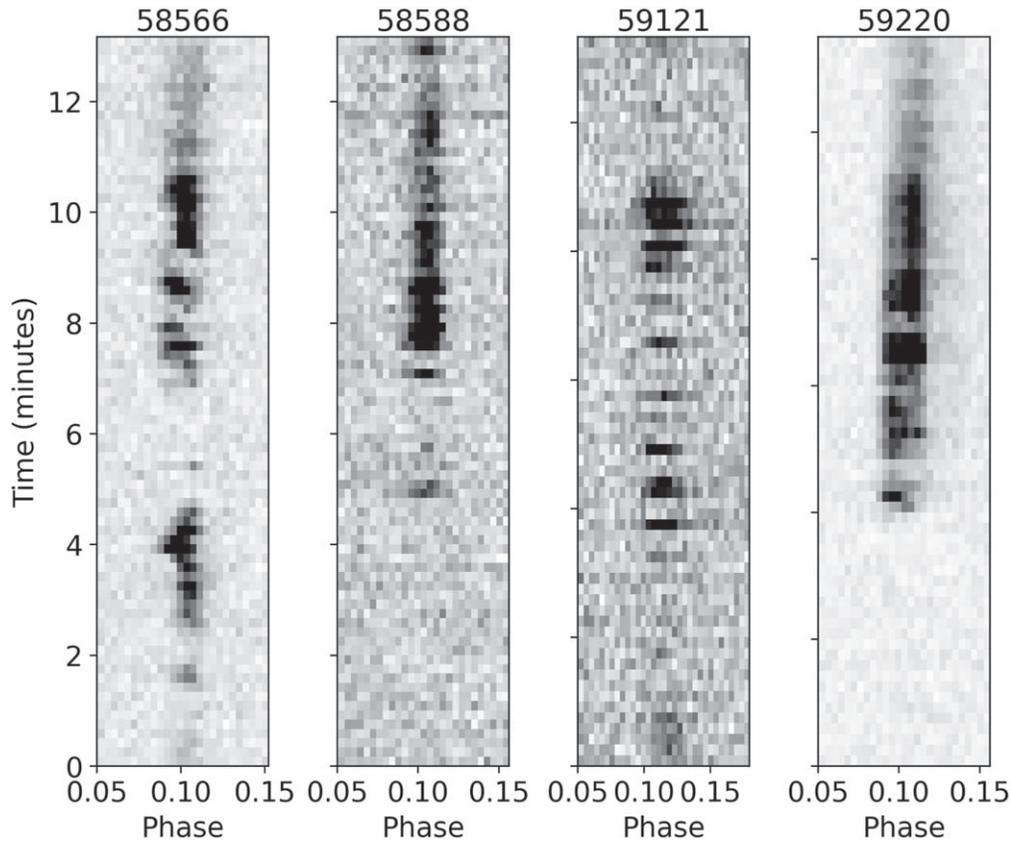

**Figure 4.** Examples of the nulling behavior in four CHIME/Pulsar observations. Each plot shows the variation in intensity of the folded pulses over time, zoomed into the on-pulse phase bins. Each time bin corresponds to 10 s. The MJD of each observation is labeled at the top of each plot.

　　timing ephemeris using the `pam` command-line utility in `psrchive` (Hotan et al. 2004; van Straten et al. 2012).[29]
2. *RFI excision.* CHIME observations are especially affected by RFI from the long-term evolution (LTE) band (spanning 700–800 MHz) in addition to numerous other terrestrial and nonterrestrial sources. We cleaned all observations of RFI using a three-step procedure. First, we applied a static RFI mask that included the most severe channels in the LTE band and other known RFI-affected channels; this mask was applied to each fold-mode archive using the `psrchive paz` command. Second, we used the `clfd` software package (Morello et al. 2019)[30] to automatically remove the remaining RFI based on the distribution of standard deviations, peak-to-peak differences, and Fourier transform amplitudes in each frequency channel and subintegration. Finally, we inspected each cleaned archive and manually removed any remaining RFI-affected channels using the `psrchive pazi` tool. In total, RFI excision typically removes ∼40% of the band.

---
[29] http://psrchive.sourceforge.net/

3. *Downsampling.* To increase the S/N of the pulse profile as a function of phase, the fold-mode data was then downsampled to 64 frequency channels (6.25 MHz resolution) and 256 phase bins and integrated completely across the observation duration. Before the temporal integration, ∼5 minutes of data acquired at the beginning and at the end of each observation were masked to avoid overweighting the low-frequency end of the transit-averaged profile.[31]

### 3.2. Periods of Nondetection and Nulling

By visual inspection, 284 of our observations yielded significant detections (i.e., the pulsar was discernible in the transit-averaged profile, with band-integrated S/N ⩾ 10), and 472 were nondetections. Figure 2 qualitatively demonstrates the pattern of CHIME/Pulsar detections and nondetections as a function of orbital motion.

PSR J2108+4516 undergoes periods of substantial eclipse, disappearing for 88–186 days (corresponding to 33%–69% of the orbit) at orbital phases roughly centered around superior conjunction. In Table 1, we have listed the MJDs encompassing major eclipses, along with their corresponding orbital phases and durations. So far, these major eclipses have occurred in the CHIME band at every observed superior conjunction, with the duration and terminating orbital phase drastically changing from orbit to orbit. In addition to these quasi-periodic disappearances, there are other, seemingly random, shorter periods of disappearance spread throughout orbital phase. These disappearances can be substantial, ranging from 19 (MJDs 59134 to 59152) to 50 days (MJDs 58624 to 58676), or of shorter duration (there are a few instances of 1–2 days disappearances, e.g., MJDs 58872 to 58873[32]).

In addition to these longer-term disappearances, PSR J2108 +4516 has exhibited instances of nulling, where pulses appear to switch on and off within a single observation. Nulling is a phenomenon exhibited by certain pulsars in which their emission suddenly disappears for one or more pulse periods (e.g., Backer 1970; Wang et al. 2007). We directly observed this behavior in only four observations, specifically MJDs 58566, 58588, 59121, and 59220. Figure 4 shows the intensity variation in these observations as a function of spin phase. The timescale over which the pulse switches on and off varies from as short as a single subintegration (10 s) to several minutes. The longest time span in which the pulsar was directly observed to be *off* occurred on MJD 58588, where the *off* state lasted at least 5 minutes (30 time bins) at the start of the observation. No

---

[30] https://github.com/v-morello/clfd
[31] The CHIME primary beam is wider at lower frequencies, and thus sensitive to pulsar transits for a longer period of time at the bottom of the CHIME band.
[32] Note: in these 1 day instances, the pulsar could disappear for anywhere from the 15 minute transit time to an entire sidereal day.

---

**Table 1**
The Encompassing MJDs, as well as Corresponding Orbital Phases and Durations for the Four PSR J2108+4516 Eclipses Observed in This Work

| Number | MJDs | Orbital Phases | Duration Days (Fraction of the Orbit) |
|---|---|---|---|
| 1 | 58422–58523 | 0.06–0.46 | 102 (38%) |
| 2 | 58685–58772 | 0.04–0.39 | 88 (33%) |
| 3 | 58956–59071 | 0.05–0.49 | 116 (43%) |
| 4 | 59237–59422 | 0.10–0.76 | 186 (69%) |

**Note.** For the orbital phases, we use the angle between the pulsar position and ascending node (i.e., the sum of the true anomaly and periastron angle; superior conjunction is at phase 0.25). Eclipse duration is given in days as well as percentage of the 269 day orbital period.

nulling behavior was seen in our filterbank observations, so no tighter constraints on the nulling timescales could be obtained.

### 3.3. Fitting of Dispersion and Scattering Variations

The extreme simultaneous DM and scattering variations exhibited by PSR J2108+4516 (as shown in Figure 3) pose a significant challenge for developing a robust timing solution. Uncorrected DM and scattering variations will decrease the accuracy and precision of individual TOA measurements and introduce chromatic noise into the resulting subbanded timing residuals (e.g., You et al. 2007; Keith et al. 2013; McKee et al. 2018). This is especially true for low-frequency broadband receivers like CHIME, where profile evolution due to interaction with the interstellar medium is significant over the band. In addition, a lack of accounting for scatter broadening will bias DM estimates across different epochs, as common methods of DM measurement will absorb some variation from the rapid changes in PSR J2108+4516's scattering properties (e.g., Demorest et al. 2013; Shapiro-Albert et al. 2021). Therefore, we needed to simultaneously estimate the scattering and DM changes for PSR J2108+4516 in order to properly assess their implications on the pulsar's local environment and to debias TOAs for robust constraints on the binary system parameters.

To this end, we used the wide-band timing methods implemented in the `PulsePortraiture`[33] software package for our analysis of PSR J2108+4516. `PulsePortraiture` allows for simultaneous fitting of TOAs, DMs, and scattering timescales using a frequency-dependent model of the pulse profile as a template (Pennucci et al. 2014; Pennucci 2019). In contrast to traditional narrow-band timing, which uses multiple TOAs across different frequencies to estimate the DM (e.g., Arzoumanian et al. 2016; Jones et al. 2017), `PulsePortraiture` encapsulates the same information in a single wide-band TOA at a select reference frequency paired with a DM measurement. Moreover, `PulsePortraiture` accounts for variable scatter broadening of the pulse profile through least-squares fitting of an exponential broadening function. In this section, we describe our `PulsePortraiture` fit setup and present our resulting DM and scattering delay estimates. We completed these estimates on the cleaned and downsampled data described in Section 3.1 (temporally integrated across the observation, with 64 frequency channels and 256 phase bins, and folded with the most up-to-date timing solution). We used the wide-band TOA and DM measurements

---

[33] https://github.com/pennucci/PulsePortraiture

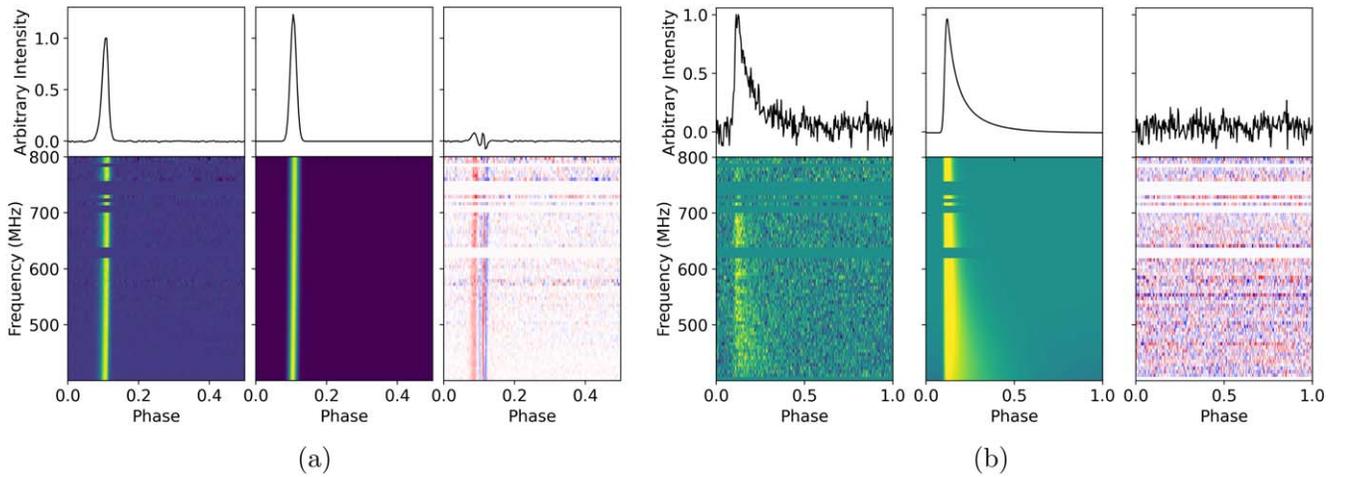

**Figure 5.** Example `PulsePortraiture` fitting results for (a) the average template of PSR J2108+4516 and (b) the pulse profile for MJD 58538, which yielded a scattering time measurement of $\tau = 3.9(2)$ ms at 1 GHz and a DM of 84.67(9) pc cm$^{-3}$. For each fit, we show the observation-integrated profile (left), the fit model (middle), and the residuals (right). The data in each of the plots are dedispersed to the fit dispersion measure.

output from this fitting for our `tempo` timing analysis in Section 3.4.

As a first step, we used the `ppalign` routine to produce a high-S/N average profile of PSR J2108+4516. This step is similar to the starting point in conventional timing analysis, where total intensity profiles are iteratively shifted to align in phase and then summed to produce an average template (Demorest 2007). However, rather than aligning the profiles with an achromatic phase shift, `ppalign` aligns pulse profiles across the band using shifts proportional to the inverse-square of the frequency (Pennucci 2019). This step corrects for variable dispersive delays between each of the summed profiles, ultimately reducing dispersive smearing in the averaged profile. Using this procedure, we summed together 14 of the highest-S/N, lowest-scattering PSR J2108+4516 profiles (as estimated from the S/N and width parameters output from the `psrchive psrstat` command). In the first iteration, we used one of the 14 profiles as a starting template for the alignment. The average profile from the initial alignment then became the template for additional alignment iterations. We completed a total of three iterations to produce our final average profile.

Next, we used the `ppgauss` routine to construct an analytic frequency-dependent model of the average profile. `ppgauss` assumes a Gaussian component decomposition of the intrinsic pulse profile, where the positions, widths, and amplitudes of the components can evolve as power-law functions with frequency (Pennucci et al. 2014). This 2D "intrinsic" Gaussian profile is then convolved with a one-sided exponential pulse broadening function (PBF), under the assumption that the scatter broadening can be modeled with the thin-screen approximation (e.g., Williamson 1972; Williamson & Scheuer 1973, 1974; McKinnon 2014). This combined model is fit to the average profile using $\chi^2$ minimization in the phase-frequency domain. For PSR J2108+4516, we modeled the intrinsic profile as a single Gaussian component and allow for the fitting of any residual scatter broadening with a fixed scattering index of $\alpha_{\rm scat} = 4.0$ (this value derives from a global analysis of 98 low Galactic latitude pulsars, yielding a frequency scaling index that is slightly shallower than the 4.4 expected from a Kolmogorov medium; Bhat et al. 2004). The resulting high-S/N average profile, fit, and residuals are shown in Figure 5(a).

The Gaussian component was fit with a full width at half maximum of $\sigma_0 = 0.01983(5)$ phase units, at $\nu_0 = 600$ MHz, which varies as a power law with the index $\alpha_\sigma = 0.12(1)$ as a function of frequency (i.e., $\sigma(\nu) = \sigma_0 (\nu/\nu_0)^{\alpha_\sigma}$). The scattering time derived from the fit was consistent with 0, indicating that the template was unscattered. The reduced $\chi^2$ was 2.2, and the residuals show a small double-peaked structure, indicating that the intrinsic profile is not perfectly Gaussian. To ensure that this does not significantly impact our DM and scattering fit accuracy, we also completed this procedure with a second Gaussian component to fit out the residual structure. We found that our DM and scattering values for the single and double Gaussian fits were consistent with each other within uncertainties. Additionally, we found that the amount of red noise in the timing residuals (see Section 5.1) was not reduced by the additional Gaussian component.

Finally, we used `pptoas` to extract simultaneous TOA, DM, and scattering time fits for each of our observation-integrated pulse profiles. For its TOA extraction functionality, `PulsePortraiture` extends the ubiquitous Fourier-domain phase-gradient shift algorithm (Taylor 1992, also known as FFTFIT) from 1D to 2D by adding a frequency dimension to the pulse template. In this extended algorithm, the fit is completed in Fourier space using nonlinear least-squares to minimize the fit statistic

$$\chi^2 = \sum_{n,k} \frac{|d_{nk} - a_n p_{nk} e^{-2\pi i k \phi_n}|^2}{\sigma_n^2}, \quad (1)$$

where $n$ indexes each frequency channel with center frequency $\nu_n$, $k$ indexes the Fourier frequency bins, $d_{nk}$ is the Fourier transform of the profile data, $p_{nk}$ is the Fourier transform of the template model, $a_n$ is the template scaling parameter, $\sigma_n$ is the data noise level, and $\phi_n$ is the phase shift applied to the template in each frequency channel. The DM of a given profile is fit by constraining the phase shifts to follow a dispersive sweep

$$\phi_n(\nu_n) = \phi_{\rm ref} + \frac{k_{\rm DM} \cdot {\rm DM}}{P_{\rm S}}(\nu_n^{-2} - \nu_{\rm ref}^{-2}), \quad (2)$$

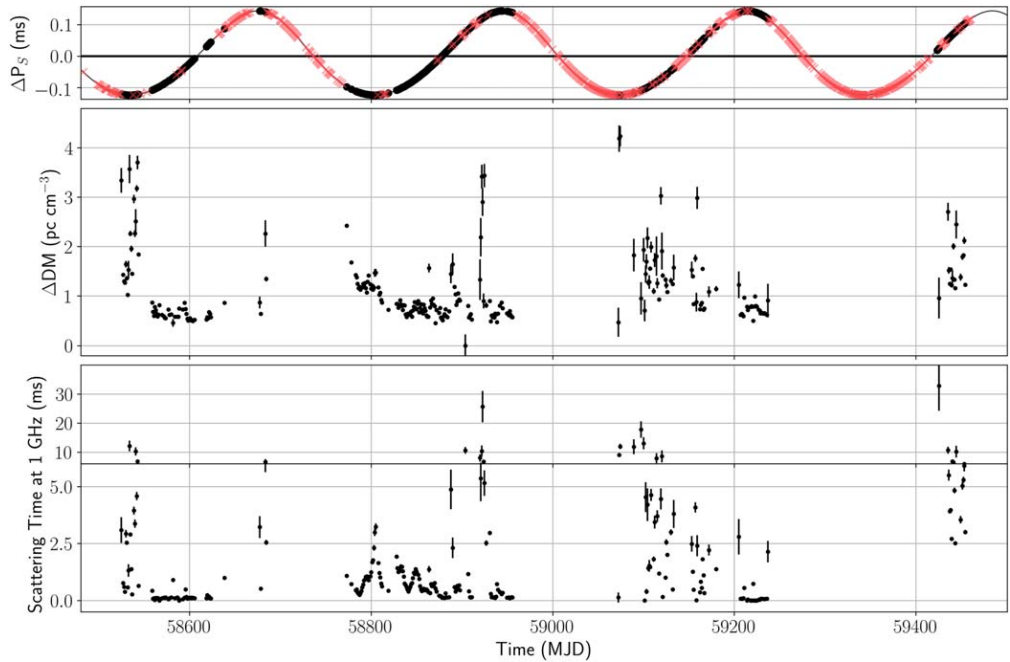

**Figure 6.** Summary of the DM and scattering variations for PSR J2108+4516 obtained from analysis of CHIME/Pulsar data. Top: the bottom panel of Figure 2 is reproduced, showing the pulsar spin period over time modulating due to binary motion. The gray line denotes the best-fit timing estimate of the orbital motion, and red x's denote epochs where timing data was acquired but no significant detection of PSR J2108+4516 was made. Middle: best-fit ΔDM values evaluated for each epoch using PulsePortraiture, showing extreme variations likely due to an inhomogeneous local environment. The ΔDM values are displayed in reference to the lowest measured DM, 81.7(2) pc cm$^{-3}$. Bottom: best-fit scattering times at 1 GHz from PulsePortraiture. The scattering axis has been split at 6 ms to enable visualization of smaller scale scattering variations. Error bars on the DM and scattering values represent $1\sigma$.

where $P_S$ is the pulsar's spin period, $k_{DM} = 0.000241^{-1}$ MHz$^2$ cm$^3$ pc$^{-1}$ s is the dispersion constant, and $\phi_{ref}$ is the phase offset at a reference frequency $\nu_{ref}$ (corresponding to the fit wideband TOA). $\nu_{ref}$ is chosen for each fit to ensure zero covariance between the TOA and DM fit values. The scattering time of a given profile is fit by convolving the ppgauss model of the average profile with a variable scattering PBF when deriving $p_{nk}$. As when fitting the average profile, we assumed a fixed scattering index of $\alpha_{scat} = 4.0$. More details about this algorithm are given in Pennucci et al. (2014), Pennucci (2019).

Using pptoas, we extracted TOAs, DMs, and scattering times for the 281 observations that yielded significant detections (as defined in Section 3.2). An example profile fit is shown in Figure 5(b) for MJD 58538. There were three MJDs (59423, 59434, 59447) with low S/Ns and residual RFI contamination for which an accurate fit could not be obtained. These MJDs are excluded from further analysis. The mean reduced $\chi^2$ of the remaining fits was ~1.2 with a standard deviation of 0.2, indicating that our model does a sufficiently good job of capturing the profile variation of the PSR J2108+4516. The resulting DM and scattering measurements are shown in Figure 6. We have scaled all of the scattering delay measurements from $\nu_{ref}$ to 1 GHz (again assuming a $\alpha_{scat} = 4.0$). The range of best-fit DMs spans from 81.7(2) to 85.9(2) pc cm$^{-3}$, a difference of 4.2(3) pc cm$^{-3}$. The range of best-fit scattering values spans from negligible to 33(9) ms. The average per-epoch TOA uncertainty was ~340 μs. We obtained $1\sigma$ DM precisions between ~$10^{-3}$ and 0.4 pc cm$^{-3}$, and $1\sigma$ scattering time precisions between ~$10^{-3}$ and ~9 ms.

The algorithm implemented by PulsePortraiture has been shown to produce reliable DM measurements (e.g., Liu et al. 2014; Guillemot et al. 2019; Nobleson et al. 2022). However, there is very limited literature where it, or a similar algorithm, has been used to simultaneously fit for scattering variations (e.g., Pennucci et al. 2014; Bilous et al. 2019; Lin et al. 2021). To ensure the accuracy of our measurements, we compared the PulsePortraiture PSR J2108+4516 output to standard techniques (such as DMX fits and frequency-resolved fitting of the scattering tails), and derived variations that were similar. In Appendix, we describe comparisons between PulsePortraiture and the profile fitting software fitburst (Masui et al. 2015; CHIME/FRB Collaboration et al. 2021), as well as fits to PsrSigSim (Shapiro-Albert et al. 2021) and PulsePortraiture simulations with similar properties to PSR J2108+4516.

Notably, MJDs 59097, 58922, and 59425 have measured scattering times of 17(3), 26(5), and 33(9) ms, respectively, and relatively low band-integrated S/N's, ranging from 14 to 20. To test PulsePortraiture's performance in this regime, we simulated data sets containing pulse profiles with high scattering values (>10 ms) and low S/Ns consistent with CHIME's noise properties (more detail provided in Figure 12 and Appendix). From these simulations, we found that the measured scattering errors are consistent with PulsePortraiture's reported uncertainties. However, there are a few noise realizations where the fit scattering value is offset from the intrinsic value by up to 10 ms. As PSR J2108+4516 is a novel source that clearly exhibits extreme scattering variations, we choose to present the scattering measurements for MJDs 59097, 58922, and 59425 as they are, but we conservatively note that the errors on these measurements may be as large as 10 ms based on our simulations.

### 3.4. Timing Analysis

We noticed significant drifts in pulse arrival times in CHIME/Pulsar data acquired using the CHIME/FRB timing parameters as an initial timing solution. To search for variations in the spin period over long timescales, we used the pdmp grid-search utility from psrchive to measure the profile drifts over pulse phase in folded CHIME/Pulsar observations and derive spin periods for each epoch. As shown in Figure 2, the resulting spin periods revealed periodic modulations in the apparent spin of PSR J2108+4516 that are typical of significant acceleration from orbital motion. As an initial characterization of this modulation, we used a weighted least-squares fitting algorithm to model the spin period variations in terms of Doppler shifts that arise from binary motion in an eccentric orbit. This algorithm yielded estimates of the intrinsic spin period ($P_S \sim 0.577$ s) as well as the five Keplerian orbital elements: the orbital period ($P_b \sim 269$ days), the orbital eccentricity ($e \sim 0.09$), the semimajor axis projected onto the plane of the sky ($x \sim 857$ lt-s), the argument of periastron ($\omega \sim 26°$), and the epoch of periastron passage ($T_0 = 58692$). Substituting these values into the Keplerian mass function assuming a pulsar mass of 1.4 $M_\odot$ and an inclination of $i = 90°$ yielded a minimum companion mass of >11 $M_\odot$. This was the first indication that the companion was a B- or O-type star.

The initial position of PSR J2108+4516, determined from the nominal pointing of the CHIME/FRB synthesized beam, has an associated uncertainty of 30′ (i.e., the beam diameter). Within this region, there are seven emission-line or Be stars, four of which have Gaia-measured distances potentially compatible with the DM distance of PSR J2108+4516 (Wenger et al. 2000; Gaia Collaboration et al. 2016, 2021). The long orbit and eclipsing nature of PSR J2108+4516 make the timing-based determination of astrometry difficult due to covariance between the binary motion, intrinsic pulsar spin-down, positional parallax, and timing variations from scattering and DM variations. We therefore proposed for observations with the Karl G. Jansky VLA, in order to identify the position of PSR J2108+4516 for timing and optical follow-up through radio imaging. Using the VLA, we successfully identified EM* UHA 138 as the optical companion of PSR J2108+4516 and fixed its position, parallax, and proper motion in our timing models to the values derived from Gaia EDR3 (shown in Table 2).[34] The details of the VLA observations and analysis are described in Section 4.

Starting from the estimated orbital parameters, we used standard narrow-band timing techniques (e.g., Alam et al. 2021) on our observation-integrated data to develop a rough initial timing solution. This initial narrow-band timing solution was then used as a starting point for high-precision modeling of the wide-band TOAs described in Section 3.3. We completed this fitting with tempo (Nice et al. 2015), with the astrometric parameters fixed to those from Gaia EDR3 and using the binary model developed by Damour & Deruelle (1986). The PulsePortraiture DM estimates and their uncertainties are added to the likelihood of the tempo generalized least-squares solver (GLS) to account for DM-related variations in TOA data (see Appendix B of Alam et al. 2021, for related discussion).[35]

The earliest observations in our data set were impacted by timing offsets occurring as a result of improper packaging of

---

[34] We note that the Gaia EDR3 position, parallax, and proper-motion values are consistent with the more recent Gaia DR3 values.
[35] This was accomplished by enabling the tempo time-variable DM model (DMX) with a bin width of 1 day, using the GLS flag when calling tempo (-G), and including DMDATA 1 in the ephemeris file.

**Table 2**
Best-fit Parameters and Derived Quantities for PSR J2108+4516

| Global Parameters | |
| --- | --- |
| Reference epoch (MJD) | 59000 |
| Observing time span (MJD) | 58526–59454 |
| Ephemeris | DE436 |
| Binary model | DD |
| Clock standard | TT (BIPM2019) |

| Timing Solution & Best-fit Metrics | | |
| --- | --- | --- |
| R.A. (J2000), $\alpha$ (h:m:s) | 21:08:23.343153 (15)[a] | ⋯ |
| Decl. (J2000), $\delta$ (d:m:s) | 45:16:24.967733 (12)[a] | ⋯ |
| Proper motion in R.A., $\mu_\alpha \cos\delta$ (mas yr$^{-1}$) | −4.291(15)[a] | ⋯ |
| Proper motion in decl., $\mu_\delta$ (mas yr$^{-1}$) | −4.912(15)[a] | ⋯ |
| Parallax, $\varpi$ (mas) | 0.306(14)[a] | ⋯ |
| Spin frequency, $\nu$ (s$^{-1}$) | 1.732408855621(3) | 1.732408858793(5) |
| First spin frequency derivative, $\dot\nu$ (10$^{-14}$ s$^{-2}$) | −1.30617(2) | −1.31127(2) |
| Second spin frequency derivative, $\ddot\nu$ (10$^{-23}$ s$^{-3}$) | | −1.709(2) |
| Spin period, $P_s$ (s) | 0.577230944506(9) | 0.577230943449(2) |
| First spin period derivative, $\dot P_s$ (10$^{-15}$ s s$^{-1}$) | 4.35211(6) | 4.36909(7) |
| Second spin period derivative, $\ddot P_s$ (10$^{-24}$ s s$^{-2}$) | | 5.696(8) |
| Orbital period, $P_b$ (days) | 269.436227(3) | 269.435003(5) |
| Orbital eccentricity, $e$ | 0.08741820(5) | 0.08742372(5) |
| Projected semimajor axis, $x$ (lt-s) | 856.26465(4) | 856.26208(5) |
| Epoch of periastron passage, $T_0$ (MJD) | 58692.83551(4) | 58692.83888(4) |
| Longitude of periastron, $\omega$ (deg) | 26.96095(5) | 26.96486(5) |
| Reduced $\chi^2$ | 2,996[b] | 90[b] |
| Number of degrees of freedom | 217 | 216 |
| Rms timing residual ($\mu$s) | 2,152 | 406 |

| Derived Quantities | | |
| --- | --- | --- |
| Spin-down luminosity, $L_s$ (10$^{33}$ erg s$^{-1}$) | 1.0 | ⋯ |
| Characteristic age, $\tau_c$ (Myr) | 2.1 | ⋯ |
| Surface magnetic field, $B_{\rm surf}$ (10$^{12}$ G) | 1.6 | ⋯ |
| Mass function, $f(M_p, M_c)$ ($M_\odot$) | 9.3 | ⋯ |
| Minimum companion mass, $M_{c,\rm min}$ ($M_\odot$) | 11.7 | ⋯ |
| 90% upper limit companion mass, $M_{c,\rm max}$ ($M_\odot$) | 113 | ⋯ |
| Galactic longitude, $l$ (deg) | 87.33996586(3)[a] | ⋯ |
| Galactic latitude, $b$ (deg) | −1.62807902(3)[a] | ⋯ |
| Parallax distance, $d_\varpi$ (kpc) | 3.26(14)[a] | ⋯ |
| DM distance (NE2001), $d_{\rm DM}$ (kpc) | 3.7 | ⋯ |

**Notes.** Dashes indicate values that are the same for both timing models. The minimum companion mass is estimated from the timing model assuming a typical pulsar mass of $M_p = 1.4$ $M_\odot$ and an edge-on orbit. Derivation of the 90% upper limit companion mass assumes that binary inclination angles are randomly distributed, such that the probability of observing a system with an inclination angle $i_0 < 26°$ is 10%.
[a] Fixed at the Gaia EDR3 values for EM* UHA 138.
[b] Note that the large reduced $\chi^2$ indicates the presence of a large amount of unmodeled noise in the data. Therefore, the parameter uncertainties from tempo reported in this table are likely underestimated.

timing data after CHIME correlator restarts. This error was fixed after MJD 58550. We removed the eight TOAs taken before this date from the analysis (MJDs 58411 to 58421). Additionally, we culled wide-band TOAs by manual inspection, roughly removing TOAs with uncertainties >500 $\mu$s. We also found a systematic achromatic timing offset between our early filterbank and fold-mode data, which was caused by a bug in the filterbank packet assembler code that was written during the commissioning of the CHIME/Pulsar instrument. We determined this offset to be exactly 251.65824 ms.[36] We added this delay to all of our filterbank TOAs to correct for this offset.

The above procedure yields the simple timing model displayed in Table 2, which includes fits to the pulsar spin period, period derivative, and the five Keplerian orbital parameters. This fit gives an rms residual of 2.2 ms and the post-fit residuals shown in the top panel of Figure 7.

There are significant variations left in the residuals, indicating that this fit may not be a complete model for the system behavior. The fit period derivative ($4.4 \times 10^{-16}$ s s$^{-1}$) implies a characteristic age of 2.1 Myr, indicating that PSR J2108+4516 is a relatively young pulsar, and thus it may exhibit significant timing noise from intrinsic irregularities in the pulsar rotation (Hobbs et al. 2010). This timing noise is traditionally modeled using the second derivative of the pulse frequency ($\ddot{\nu}$; Lyne 1999). Pulsar/massive-star binaries have also exhibited other post-Keplerian effects such as changes in the orbital period ($\dot{P}_b$), longitude of periastron ($\dot{\omega}$), or projected semimajor axis ($\dot{x}$). In an attempt to characterize the remaining variation in the PSR J2108+4516 residuals, we completed three more fits to the wide-band TOAs with the following additional parameters: (1) $\ddot{\nu}$ to model the long-term behavior of the timing noise; (2) $\ddot{\nu}$ and $\dot{P}_b$ to account for any changes in the orbital period; and (3) $\ddot{\nu}$, $\dot{\omega}$, and $\dot{x}$ to try and detect spin–orbit coupling or other kinematic effects.

As expected, fitting for $\ddot{\nu}$ significantly improved the fit, reducing the rms residual to 406 $\mu$s and the reduced $\chi^2$ from 2,996 to 90. The resulting post-fit residuals are shown in the bottom panel of Figure 7, and the timing parameters are shown in the right-most column of Table 2. The fit value for the second derivative of the frequency was $\ddot{\nu} = -1.709(2) \times 10^{-23}$ s$^{-3}$ (a second period derivative of $\ddot{P} = 5.696(8) \times 10^{-24}$ s s$^{-2}$). We discuss the implications of this fit in Section 5.1.

Simultaneously fitting for both $\ddot{\nu}$ and $\dot{P}_b$ did not significantly improve the fit, yielding the same rms residual and a reduced $\chi^2$ as the fit with only $\ddot{\nu}$. In addition, simultaneously fitting for $\ddot{\nu}$, $\dot{\omega}$, and $\dot{x}$ did not result in a significantly better fit.

## 4. VLA Observations and Localization

To accurately localize PSR J2108+4516 and further confirm the companion, we observed the field around the best available narrow-band timing position of PSR J2108+4516 for 2 hr using the VLA on 2020 May 19 (MJD 58988).[37] The scheduled observation date coincided with an expected eclipse based on the contemporaneous CHIME data set, given that neither CHIME/Pulsar nor CHIME/FRB had detected significant pulsations since 2020 April 17 (MJD 58956). However, we believed the long eclipsing time (∼100 days) was caused by interaction of the radio signal with circumstellar material and/or winds given the dramatic electromagnetic variations mentioned in Section 3. We decided to observe the source with the VLA at S band (2–4 GHz, using 2,048 frequency channels) as any significant pulsar emission is expected to be scattered less than that in the CHIME band.

Raw visibility data were recorded in the VLA "pulsar" mode,[38] where visibilities are coherently integrated into 20 pulse phase bins modulo the initial timing model obtained from CHIME/Pulsar data. After applying the standard processing routines from the Common Astronomy Software Applications (CASA; McMullin et al. 2007) pipeline, we first searched for radio point sources in the observed field by integrating phase-binned visibilities over the entire observation, using the Briggs *robust* weighting scheme to create the cleaned, composite image. With this approach, we detected a radio source with the VLA at the position of EM* UHA 138 (Gaia Collaboration et al. 2021) in the cleaned image. Figure 8 shows the composite image and a zoomed-in view of the radio source coincident with EM* UHA 138.

In order to assess the temporal nature of the radio source, we created images for each phase-binned visibility data set. The phase-binned images are also shown in Figure 8 below the composite images. The radio source is only present in two distinct phase bins, which is consistent with a strong radio pulsation at the pulsar rotation period. Based on the consistency in timing, we concluded that the radio source is PSR J2108+4516.

## 5. Discussion

The radio-timing observations of PSR J2108+4516 establish that the source orbits a massive star, with, given the large variations in DM and scattering time, a clumpy circumstellar medium. Moreover, the VLA localization makes it clear that the companion is EM* UHA 138, a bright star for which multiple archival observations are available. In this section, we interpret radio-timing and localization results in the context of the association with EM* UHA 138.

### 5.1. Timing Implications

The best-fit timing parameters (Table 2) establish that PSR J2108+4516 is a 0.577 s pulsar in a 269 day orbit of low eccentricity ($e = 0.09$) with a spin-down luminosity of $1.0 \times 10^{33}$ erg s$^{-1}$, a characteristic age of 2.1 Myr, and a surface magnetic field of $1.6 \times 10^{12}$ G.

When fitting only the pulsar spin period, period derivative, and the five Keplerian orbital parameters, the timing solution still contains significant variations in the residuals. We found that these variations can be reduced the most by fitting a second derivative of the spin period $\ddot{P} = 5.696(8) \times 10^{-24}$ s s$^{-2}$ with a corresponding first period derivative of $\dot{P} = 4.36909(7) \times 10^{-15}$ s s$^{-1}$. This second derivative value far exceeds that expected from standard pulsar spin evolution caused by magnetic braking (assuming a dipole braking index of $n = 3$ yields an expected value of $\ddot{P} = 10^{-30}$ s s$^{-2}$ for PSR J2108

---

[36] More specifically, after coherent dedispersion, a fixed number of data samples are thrown out to remove edge artifacts from cyclic convolution (Naidu et al. 2015). The code that determines the metadata timestamps for the filterbank packets did not take this data removal into account, resulting in a constant offset corresponding to the number of data samples removed. Thus, this offset value is fixed and exact.

[37] These observations were acquired as part of Director's Discretionary Time project 20A-474.

[38] https://science.nrao.edu/facilities/vla/docs/manuals/oss/performance/pulsar

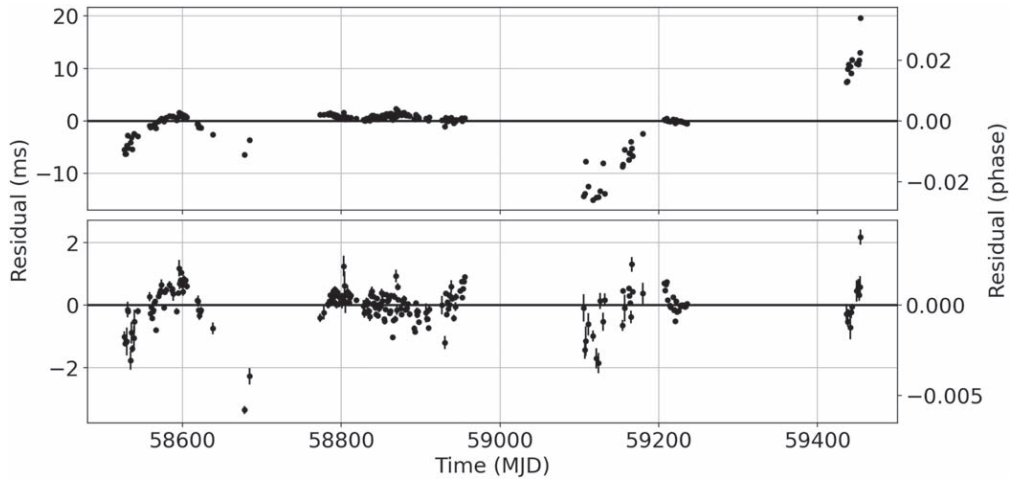

**Figure 7.** Post-fit residuals from a fit to the `PulsePortraiture` TOAs for (top) spin period, period derivative, and the five Keplerian orbital parameters and (bottom) the same model plus a second derivative of the spin period. The left axis shows the residual amplitude in milliseconds; the right axis shows the residual amplitude in spin period phase.

+4516). One possible explanation for such a significant $\ddot{P}$ is red timing noise from intrinsic imperfections in the pulsar rotation rate. This type of timing noise is more prominent in young pulsars with large period derivatives (Arzoumanian et al. 1994). To determine if the magnitude of timing noise exhibited by PSR J2108+4516 is consistent with that expected intrinsically for our measured $\dot{P}$, we use the $\Delta_8 = \log\left(\frac{1}{6\nu}|\ddot{\nu}|T_8^3\right)$ metric developed by Arzoumanian et al. (1994), which quantifies the timing noise strength based on the value of the cubic term $\ddot{\nu}$ fit over an observing span of $T_8 = 10^8$ s. Using values from our cubic fit in base units of seconds, we calculate $\Delta_8 = 0.22$, whereas the metric we expect for pulsars of the same $\dot{P}$ is $\Delta_8 = -2.0$ (see the relation given in Equation (3) and Figure 1 of Arzoumanian et al. 1994). Thus, the timing noise present in the PSR J2108+4516 residuals is 2–3 orders of magnitude stronger than the typical value we expect to occur purely from intrinsic irregularities in the pulsar rotation.[39] This suggests that at least some of the timing noise is not intrinsic to the pulsar.

The origin of this additional timing noise remains uncertain. In such a complicated system, there are possible physical explanations, such as interactions between the companion and surrounding material that affect the pulsar's spin. Future timing at higher frequencies, where the pulsar signal is less dispersed and scattered, will help reveal the nature of this noise.

### 5.2. High-mass Companion

The VLA localization presented in Section 4 allows us to confidently associate PSR J2108+4516 with the O/Be-type star EM* UHA 138 (e.g., Merrill & Burwell 1950; Hardorp et al. 1964; Wackerling 1970; Kohoutek & Wehmeyer 1997), located in the North America nebula of the Cygnus region. In this section, we review the optical archival observations associated with EM* UHA 138 in combination with mass constraints from the timing solution.

#### 5.2.1. Brightness and Astrometric Parameters

EM* UHA 138 has an estimated $V$ magnitude $m_V \sim 11$ mag with strong H$\alpha$ emission lines, and is historically noted as being variable (Welin 1973). The Gaia EDR3 catalog (Gaia Collaboration et al. 2016, 2021) entry[40] for EM* UHA 138 provides updated measurements of the optical companion position, proper motion, and parallax, which we have listed in Table 2. The measured parallax of $\varpi = 0.306(14)$ mas corresponds to a distance of $d_\varpi = 3.26(14)$ kpc. EM* UHA 138 has a relatively large excess astrometric noise, $\epsilon = 0.112$ mas, measured with 16$\sigma$ significance, which is a scatter term that captures the systematic errors and additional intrinsic motion beyond what is parameterized in the astrometry model. In this system, $\epsilon$ is likely dominated by the binary motion, i.e., the semimajor axis for the pulsar is $a_p \sim 2$ au, and if we assume a mass ratio $q = M_p/M_c \sim 0.1$ ($M_p$ and $M_c$ are the pulsar and companion mass, respectively), then the semimajor axis for the companion is $a_c \sim 0.2$ au, which corresponds to an astrometric wobble $a_c/d_\varpi \sim 0.1$ mas.[41]

The Gaia calibration-corrected G magnitude of EM* UHA 138 is $m_G = 10.976(4)$ mag. We also inspected the ASAS-SN $V$-band photometry data via the ASAS-SN Photometry Database[42] (Kochanek et al. 2017), which reports a mean V magnitude of $m_V = 10.79(2)$ mag, in line with the Gaia measurements and previous literature values.

---

[39] As a caveat, Shannon & Cordes (2010) note that the $\Delta_8$ metric assumes a total observing span of $T_8 = 10^8$ s = 3.2 yr, and that it should not be used to compare pulsars with different observation lengths as $\ddot{\nu}$ increases as a function of $T$. Instead, they propose a $\sigma_{TN}$ metric based on the rms of the residuals and the average TOA error after fitting for $\nu$ and $\dot{\nu}$ only. They then empirically derive the expected a scaling law relating $\sigma_{TN}$ to $\nu$, $\dot{\nu}$, and $T$ through a maximum likelihood fit to a large sample of pulsars. Although our observing span of 2.8 yr is not drastically different than that assumed for $\Delta_8$, as a sanity check we calculate the Shannon & Cordes (2010) parameter to be $\ln(\sigma_{TN}) = 7.7$, with $\sigma_{TN}$ in units of $\mu s$. Meanwhile, the expected relationship that Shannon & Cordes (2010) derived for "canonical" pulsars (i.e., not millisecond pulsars or magnetars) predicts $\ln(\hat{\sigma}_{TN}) = 6.0$ with a $\sigma = 1.6$ dex scatter. Thus, by this metric, the timing noise exhibited by PSR J2108+4516 is 1$\sigma$ higher than what is expected.

[40] Unique identifier: Gaia EDR3 2162555482829978496.
[41] Note that the astrometric precision of Gaia DR4 is predicted to be 0.01 – 0.02 mas for for stars brighter than $G = 15$ magnitude (see the Extended Science Performance for the Nominal and Extended Mission at www.cosmos.esa.int/web/gaia/science-performance). Combined with timing observations, this precision will allow us to obtain a full astrometric solution for the system.
[42] APJ210823.34+451624.9: https://asas-sn.osu.edu/photometry/18e4ae56-e41e-5dbe-b619-655140297bcc.

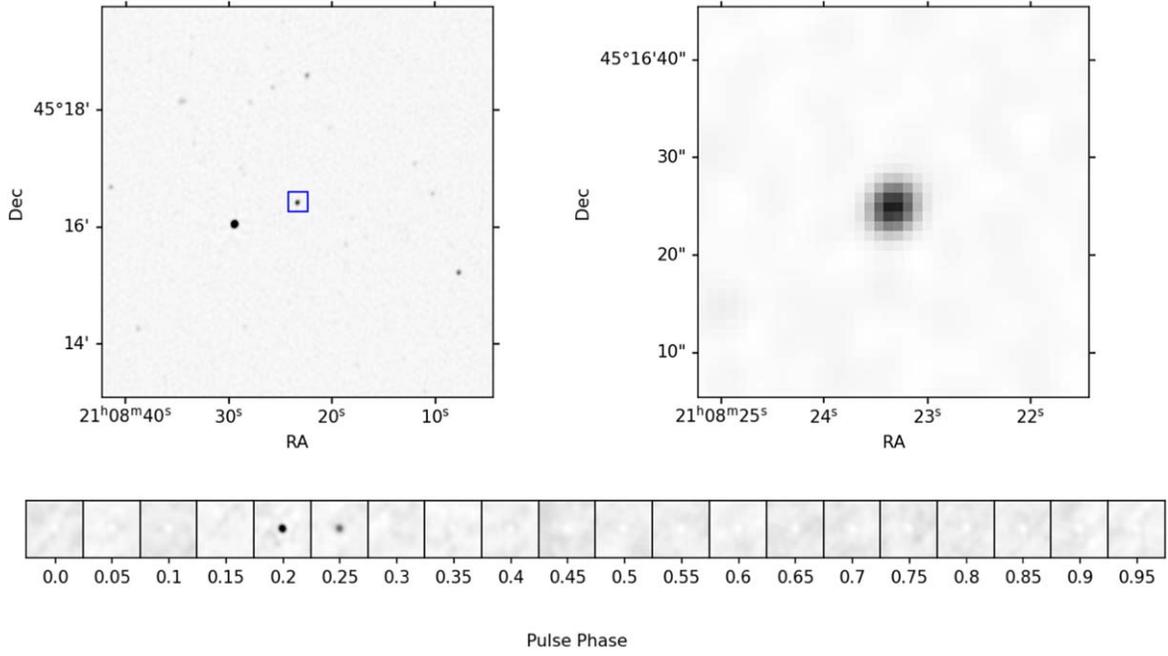

**Figure 8.** VLA continuum image of the sky field around PSR J2108+4516 at 2 to 4 GHz. The top row of images were created by integrating 1.7 hr of pulse phase-averaged visibility data. The upper left panel shows a zoomed-out image of the sky field. The blue box highlights the emission at R.A. $21^h08^m23\overset{s}{.}343$ and decl. $+45°16'24\overset{''}{.}9657$, the location of EM* UHA 138. The upper right panel is a zoomed-in image of the blue box region on the left. The lower panel shows continuum emission as a function of pulse phase. Since the radio emission is only detected at pulse phases 0.2 and 0.25, we confirm that the radio emission is from PSR J2108+4516, and, given the positional coincidence, that the companion is indeed EM* UHA 138. Note that the detection of the source in both the 0.2 and 0.25 bins could be consistent with large changes in the DM or scattering of the profile during eclipse.

### 5.2.2. Mass Constraints

From the radio-timing analysis, we are able to place upper and lower bounds on the mass of EM* UHA 138. The mass function of the binary system can be computed from the orbital period and the projected semimajor axis via

$$f(M_p, M_c) = \frac{(M_c \sin i)^3}{(M_p + M_c)^2} = \frac{4\pi^2}{T_\odot} \frac{x^3}{P_b^2}, \quad (3)$$

where $T_\odot = GM_\odot c^{-3} = 4.925490947641267$ $\mu$s is an exact constant describing the light travel time across a $1\,M_\odot$ gravitational radius, and $i$ is the orbital inclination angle of the binary system. For the PSR J2108+4516 system, we measure $f(M_p, M_c) \approx 9.3\,M_\odot$. By assuming a typical pulsar mass, $M_p = 1.4\,M_\odot$, and an edge-on orbit, $i = 90°$, we can solve Equation (3) for a lower bound on the companion mass: $M_{c,\min} \sim 11.7\,M_\odot$. With the additional assumption that the orbital inclination angles are randomly distributed, where $\Pr(i < i_0) \sim 1 - \cos i_0$, we can estimate a 90% upper limit by asserting $i_0 = 26°$, producing $m_{c,\max} \sim 113\,M_\odot$. However, we note that such a shallow inclination angle is very unlikely given the substantial eclipses exhibited by PSR J2108+4516.

In addition to the timing limits, we can use estimates of the absolute magnitude of EM* UHA 138 to determine its spectral type and a range of plausible masses. The parallax distance of $d_\varpi \approx 3.3$ kpc corresponds to a distance modulus $\mu = 12.6(1)$ mag, where we note that there is a substantial jump in the reddening likely due to the star-forming region associated with the North America nebula (Green et al. 2019).[43] The range in interstellar reddening at this sky location, $0.62 < E_{g-r} < 0.82$ mag, corresponds to an extinction $A_V \approx 3.68 E_{g-r} \approx 2.6(4)$ mag (Schlegel et al. 1998). Therefore, the absolute magnitude of EM* UHA 138 is $M_V = m_V - \mu - A_V \approx -4.5(5)$ mag (where we have adopted the ASAS-SN V-band apparent magnitude, $m_V = 10.79(2)$ mag). Assuming EM* UHA 138 is on the main sequence, the estimated absolute magnitude implies a spectral type (e.g., Pecaut & Mamajek 2013)[44] in the range B0Ve ($M \sim 17.5\,M_\odot$, $R \sim 7R_\odot$, $T_{\text{eff}} \sim 30,000$ K) to O8Ve ($M \sim 23\,M_\odot$, $R \sim 9R_\odot$, $T_{\text{eff}} \sim 36,000$ K), consistent with the lower-mass limit calculated from the pulsar timing data. The nominal Gaia Data Release 3 (DR3) GSP-Phot effective temperature is $3.64(2) \times 10^4$ K, on the upper end of this range. However, this value could be uncertain by up to 1000 K for sources with high extinction in the Galactic plane (see comparison to APOGEE DR16 values in Table 1 and Figure 8 of Andrae et al. 2022). Acquiring a high-quality optical spectrum of EM* UHA 138 will provide better constraints on the temperatures and, therefore, the inferred mass and radius.

If we take the range of potential companion masses, assume a typical pulsar mass $M_p = 1.4\,M_\odot$, and use the measured mass-function $f(M_p, M_c) \approx 9.3\,M_\odot$, we can reevaluate Equation (3) to determine a range of possible orbital inclination angles. This mass range $17.5\,M_\odot < M_c < 23\,M_\odot$ corresponds to a inclination angle $50°.3 \lesssim i \lesssim 58°.3$, which is insufficiently inclined to yield edge-on *hard* eclipses (where the companion star directly blocks pulsations from PSR J2108+4516).

---

[43] See, e.g., the 3D Dust Mapping with Pan-STARRS online tool, http://argonaut.skymaps.info/.

[44] Spectral types and parameters retrieved from Eric Mamajek's "A Modern Mean Dwarf Stellar Color and Effective Temperature Sequence," http://www.pas.rochester.edu/~emamajek/EEM_dwarf_UBVIJHK_colors_Teff.txt.

### 5.3. DM Variations and Local Environment

Figure 6 clearly demonstrates that the dispersion and scattering properties of PSR J2108+4516 show extreme variations on intra-day timescales over the course of a single orbit. Over the entire observing period, the observed DMs range from 81.7(2) to 85.9(2) pc cm$^{-3}$, a difference of 4.2(3) pc cm$^{-3}$, and the observed scattering times at 1 GHz range from negligible to 33(9) ms. The largest day-to-day change in DM was 3.7(4) pc cm$^{-3}$ (from MJD 59072 to 59073), and the largest day-to-day change in scattering was 19(5) ms (from MJD 58922 to 58923).

PSR J2108+4516's day-to-day DM variations exceed the amplitude of variations seen in most of the pulsar population over timescales of years (e.g., Petroff et al. 2013). We can use the maximum ΔDM measurement to estimate the variation in electronic content in the PSR J2108+4516 system. To accomplish this, we model total DM in terms of an ISM and orbital component, i.e., $DM_{tot} = DM_{ISM} + DM_{orb}$, where

$$DM_{orb} = \int_r^x n_{e,orb}(r') dr'; \quad (4)$$

$-x < r < x$ is the projected distance of the pulsar from the plane of the sky, and $x \sim 850$ lt-s is the projected semimajor axis of the orbit. We assume that the large intra-day DM variations observed from PSR J2108+4516 arise from an evolving $DM_{orb}$, as typical values of $|d(DM_{ISM})/dt|$ are on the order of $\sim 0.01$ pc cm$^{-3}$ yr$^{-1}$ (e.g., Petroff et al. 2013). The observed variation can therefore be related to the change in electron number density local to the binary system, because $\Delta DM \sim \Delta DM_{orb} \approx n_{e,orb} \Delta r$. With $\Delta r = 2x \approx 1700$ lt-s, we find that the local electron number density varies by at least $n_{e,orb} = \Delta DM_{orb}/(2x) \sim 10^5$ cm$^{-3}$.

The pattern of nondetections (Section 3.2) and DM–scattering variations (Figure 6) exhibited by PSR J2108+4516 suggests a complex and dynamic surrounding environment influenced by a disk/wind from the companion star as well as possible local structure in the encompassing HII region. The signal is completely eclipsed in the CHIME band for 33%–69% of the orbit roughly aligned with the pulsar passing behind the companion. A hard eclipse is not likely possible, because the constraints on the companion mass and spectral type predict orbital inclinations on the order of $\sim 50°$ (as derived in Section 5.2.2). Instead, this pattern of nondetection around superior conjunction is likely due to smearing, scattering, and/or absorption of the pulsations by material in a circumstellar disk surrounding the companion or a very dense stellar wind. This scenario is bolstered by our VLA detection of the pulsar at S band (2–4 GHz) while it was simultaneously obscured in the CHIME band (400–800 MHz), suggesting that the obscuration mechanism is frequency-dependent.

Shorter periods of nondetection outside of eclipse are an indication that the disk/wind is anisotropic and clumpy. This is supported by the scattering and DM measurements, which show drastic variations and structure even when the pulsar is nominally in front of the companion. Notably, from MJD 58773 to 58955, the scattering measurements show quasi-periodic oscillations of variable amplitude with a period on the order of 20 days (see Figure 6). Another short span from MJD 58911 to 58932 includes a significant number of nondetections as well as high scattering measurements up to 26(5) ms. The disk/wind structure is also clearly highly variable, as the phases of apparent eclipse egress change dramatically from orbit to orbit (as summarized in Table 1). This variability is again reflected in the DM–scattering measurements, which change significantly from day to day and orbit to orbit. Such variation in the DM from orbit to orbit has also been observed in pulsar/massive-star binary J1740-3052 (Madsen et al. 2012). In the context of this chaotic environment, it is possible that the four instances of nulling we observed were caused by interactions between the pulsar and rapidly moving ionized material. If so, one might expect frequency-dependent nulling, which may be investigated in a future work.

Another notable feature is the difference between the DM variations on either side of eclipse. During ingress the pulsar disappears abruptly while during egress there is a more gradual downward *swoop*. This is illustrated in Figure 9, which shows DM variations as a function of orbital phase for the four orbits that we have observed with CHIME/Pulsar. This asymmetry across the eclipse could be a symptom of the underlying geometry of the system. For example, in Johnston et al. (1996), Melatos et al. (1995), it was proposed that a similar asymmetry about the eclipse of PSR B1259−63 could be explained by a circumstellar disk combined with the large eccentricity ($e \sim 0.9$) of the pulsar orbit and misalignment of the major axis of the orbit with the line of sight (the longitude of periastron, $\omega$, is not equal to 90°, so that the pulsar spends more time behind the disk on one side of eclipse). In Figure 9, we plot an example of the DM variations expected from different configurations of a Be star disk in the PSR J2108+4516 system. We use a toy model for the disk given in Melatos et al. (1995), which models the disk density radial fall-off as an inverse quadratic. We manually select the parameters for this model that best match the PSR J2108+4516 DM measurements while falling within the typical range given in Be star disk literature (e.g., Waters et al. 1991; Bjorkman 1997; Klement et al. 2017). Since the eccentricity of PSR J2108+4516 is significantly lower than PSR B1259−63, it was not possible to reproduce the asymmetry in the DM variations only using misalignment of the major axis of the orbit with respect to the line of sight. Instead, in our model, producing asymmetry required inclination and misalignment of the disk itself. More specifically, the asymmetric model (represented by the solid black line in Figure 9) assumes an orbital inclination of 60°, with the Be star disk inclined an additional 30° with respect to the pulsar orbital plane and rotated by 40° with respect to the major axis of the orbit ($\theta_d$ and $\omega_d$ in Figure 8 of Melatos et al. 1995). Given the large orbit-to-orbit variations in the DM structure, it is likely that our simple disk model is not a complete explanation of the DM variations observed in PSR J2108+4516. However, overall, this DM asymmetry around eclipse supports the presence of some inclination of both the pulsar orbit and a disk-like structure, which is plausible if there is a *kick* imparted during the pulsar birth. Similar spin–orbit misalignment angles of $\sim 20°$–30° have been seen in the pulsar/massive-star binary systems PSR J0045−7319 and PSR B1259−63, with implied kick velocities $\sim 100$–200 km s$^{-1}$ (Kaspi et al. 1996; Shannon et al. 2014). However, these systems have much higher eccentricities ($e \sim 0.80$ and 0.87 for PSR J0045−7319 and PSR B1259−63, respectively) than PSR J2108+4516 ($e \sim 0.09$), in which case the $\sim 30°$ misalignment for PSR J2108+4516 could be explained by a lower magnitude kick, particularly if fortuitously directed. Alternatively, a strikingly similar asymmetry about eclipse ingress and egress observed in some black widow pulsar systems has been attributed to a comet-like "tail" of material being swept back from the companion by orbital motion (e.g., Fruchter et al. 1990; Polzin et al. 2018). More detailed modeling of the PSR J2108+4516 DM variations during ingress and egress could help constrain the properties of the surrounding disk/wind as well as the geometry of the orbit.

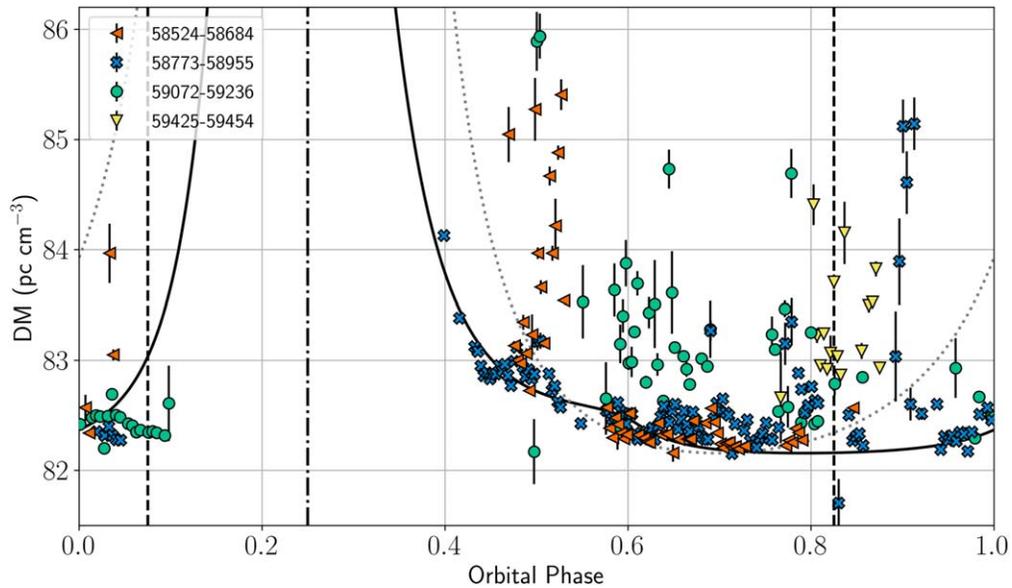

**Figure 9.** DM variations folded as a function of orbital phase (the calculation of phase here is consistent with Table 1). Each marker represents DM variation from a different orbit. The vertical dashed–dotted line indicates superior conjunction (orbital phase of 0.25). The vertical dashed line represents the phase of periastron. The dashed gray line indicates the DM variation expected from a typical Be star disk assuming both the disk and pulsar orbit are edge-on with respect to the line of sight. In this model, we assume a companion star radius of $R_c = 7\ R_\odot$, and a central electron density of $n_{e,0} = 10^{12}$ cm$^{-3}$. Radially, the disk density falls off as $n_e(r) = n_{e,0}(r/R_c)^{-n}$, with $n = 3.8$, and a hard cutoff at $R_{\mathrm{out}} = 100 R_c$; vertically, the density falls off with a scale height $0.1 R_{\mathrm{out}}$. The solid black line indicates the same disk model, but assuming a more complicated alignment of the disk and the pulsar orbit to achieve asymmetry in the DM variations during eclipse ingress and egress. Modeled DM variations are normalized to the lowest DM observed from PSR J2108+4516.

### 5.4. Accretion Constraints

Despite the large DM variations being indicative of highly dense material surrounding the system, we anticipate that accretion onto the pulsar is likely not actively occurring. Initial support for this assumption comes from the fact that phase-coherent timing of the pulsar has been achieved over many orbits. If significant mass transfer were occurring, the pulsar would be subject to accretion torques that would significantly vary the pulse period, making phase-coherent timing challenging (e.g., Bildsten et al. 1997).

Additionally, our timing constraints indicate that EM* UHA 138 is not filling its Roche lobe. Assuming a $1.4\ M_\odot$ pulsar, we derive a range of mass ratios $q$, where $q = 0.12$ for the lower-limit companion mass ($M_{c,\mathrm{min}} \sim 11.7\ M_\odot$), and $q = 0.012$ for the upper limit companion mass ($M_{c,\mathrm{min}} \sim 113\ M_\odot$) obtained from timing. Using Eggleton's approximation (Eggleton 1983), we find that the ratio between the Roche lobe radius and the orbital separation ($a$, where $a = x/\sin(i)$) ranges from 0.22 ($q = 0.12$) to 0.11 ($q = 0.012$). Given that $a\sin(i) = 856$ lt-s, and taking $i = 90°$ for the minimum mass, and $i = 26°$ for the maximum mass, we find corresponding Roche lobe radii of $80\ R_\odot$ and $92\ R_\odot$, respectively. As the radius of an $11.7\ M_\odot$ main-sequence star is $\sim 5\ R_\odot$, and the radius of an $113\ M_\odot$ main-sequence star is $\sim 12\ R_\odot$ (Demircan

& Kahraman 1991), in both cases the companion star lies comfortably inside the Roche radius (even in the case that the companion is an OB supergiant, typical radii range from 20 to $30\ R_\odot$; e.g., Remie & Lamers 1982). Thus, efficient mass transfer through Roche lobe overflow is unlikely.

However, Roche lobe overflow is not the only possible route for mass transfer in this system. The DM variations observed in PSR J2108+4516 combined with the strong Hα emission lines in EM* UHA 138's spectrum imply the presence of a circumstellar disk or very dense stellar wind, which may accrete onto the pulsar. This is the case for all Be/X-ray binaries, which accrete from a circumstellar disk, and some supergiant/X-ray binaries, which accrete from a dense stellar wind. In this scenario, there are two major barriers that the disk/wind matter must overcome in order to accrete onto the pulsar surface: the *pulsar radiation barrier* and the *magnetospheric boundary*. We examine each of these barriers in the context of the disk model derived in Section 5.3.

Assuming a uniform medium moving at velocity $v_{\mathrm{rel}}$ relative to the pulsar, classical Bondi–Hoyle–Lyttleton accretion theory predicts that all matter within a radius of $R_{\mathrm{acc}} = 2GM_p/v_{\mathrm{rel}}^2$ will be accreted (where $M_p$ is the mass of the pulsar; Hoyle & Lyttleton 1939). The accretion rate is given by $\dot{M}_{\mathrm{acc}} = \rho v_{\mathrm{rel}} \pi R_{\mathrm{acc}}^2$, where $\rho$ is the density of the medium. If the relativistic pulsar wind is strong enough, it will halt accretion by forming a shock with the companion wind that is outside the accretion radius of the pulsar. Campana et al. (1995) derives the minimum mass capture rate $\dot{M}_{\mathrm{prb}}$ needed to overcome this "pulsar radiation barrier" by equating the pulsar radiation pressure with the companion wind ram pressure at the accretion radius:

$$\dot{M}_{\mathrm{prb}} = \frac{L_{\mathrm{sd}}}{c v_{\mathrm{rel}}}, \qquad (5)$$

where $c$ is the speed of light, and $L_{sd}$ is the spin-down luminosity of the pulsar. To evaluate the efficacy of this barrier for the PSR J2108+4516 system, we estimate the relative velocity using $v_{rel} = [(v_x - v_\phi)^2 + v_r^2]^{1/2}$ where $v_x$ is the orbital velocity of the pulsar, and $v_\phi$ and $v_r$ are the rotational and radial velocities of the companion disk, respectively (King & Cominsky [1994](#)). The orbital velocity is given by $v_x = GM_c(2/r - 1/a)$, where $M_c = 20\ M_\odot$ is the mass of the companion, $r$ is the radial distance of the pulsar from the companion, and $a \sim 856\ \text{lt-s}/\sin(60°) = 988$ lt-s is the semimajor axis of the orbit (we note that assuming an orbital inclination of 50°.3, the lower end of the inclinations implied from optical constraints in Section [5.2.2](#), does not significantly change the resulting calculations). Following the Be star disk wind model summarized by Waters & van Kerkwijk ([1989](#)), we take the radial velocity of the wind to be $v_r(r) = v_{r0}(r/R_c)^{n-2}$, with a corresponding density distribution of $\rho(r) = \rho_0(r/R_c)^{-n}$, where $R_c = 7\ R_\odot$ is the companion equatorial radius, $\rho_0 = 1.6 \cdot 10^{-12}$ g cm$^{-3}$ is the central density of the disk, and $n = 3.8$ (note: to simplify our calculations, we ignore the possible inclination of the disk with respect to the pulsar orbit that was proposed in Section [5.3](#)). As highlighted by Waters & van Kerkwijk ([1989](#)), $v_{r0}$ is not very well constrained, but is likely less than the sound speed in the wind ($\sim$10 km s$^{-1}$ for a typical stellar wind). We adopt a value of $v_{r0} = 5$ km s$^{-1}$. Finally, we take the rotational velocity of the disk to be $v_\phi(r) = v_{\phi 0}(r/R_c)^{-1}$, where $v_{\phi 0} = 600$ km s$^{-1}$ is the rotational velocity of the Be star (which is $\sim$80% of the critical break-up velocity; Townsend et al. [2004](#)).

Combining the framework described above with timing results from Table [2](#), we calculate the pulsar radiation barrier limit to be $\dot{M}_{prb} \sim 10^{-14}\ M_\odot\ \text{yr}^{-1}$ at both apastron ($r = a(1+e)$) and periastron ($r = a(1-e)$), while the expected Bondi–Hoyle–Lyttleton accretion rate is only $\dot{M}_{acc} \sim 10^{-18}\ M_\odot\ \text{yr}^{-1}$. Thus, for the disk model derived in Section [5.3](#), we do not expect the pulsar radiation barrier to be overcome. However, as this model was manually fit by eye, the resulting parameters are highly uncertain. If, for example, the density profile index is flattened to $n = 2$, which is on the lower end of the possible range for Be star disks, the accretion rate at periastron increases to $\dot{M}_{acc} \sim 10^{-14}\ M_\odot\ \text{yr}^{-1}$. Thus, we do not fully rule out the possibility that mass from EM* UHA 138's wind could overcome the pulsar radiation barrier.

The next barrier that the disk wind needs to overcome is the *magnetospheric boundary*. At a specific magnetospheric radius, the motion of the infalling gas becomes dominated by the magnetic field of the pulsar, such that the matter is forced to corotate with the neutron star. If the pulsar is spinning too fast, then the corotation velocity will exceed the Keplerian velocity, and the infalling material will be spun away from the neutron star rather than accreting onto the surface (Illarionov & Sunyaev [1975](#)). The spin period for which the Keplerian and corotation velocity are balanced at the magnetospheric radius is called the "equilibrium spin" (Waters & van Kerkwijk [1989](#)). Using Equations (13), (14), and (15) in Waters & van Kerkwijk ([1989](#)), we find that the equilibrium spin for PSR J2108+4516 is $P_{eq} \sim 10^3$ s at both periastron and apastron. This is several orders of magnitude higher than PSR J2108+4516's actual spin period of 0.577 s. Even when relaxing the density profile index to $n = 2$, the equilibrium spin only lowers to $P_{eq} \sim 10^2$ s. Thus, we conclude that the magnetospheric boundary is not overcome in the PSR J2108+4516 system under the assumption of a typical Be star disk wind, and thus, accretion onto the neutron star surface is very unlikely.

### 5.5. Origin and Evolution

Of the five previously known radio pulsar/high-mass companion binary systems for which orbital eccentricity has been measured (Johnston et al. [1992](#); Kaspi et al. [1994](#); Stairs et al. [2001](#); Lyne et al. [2015](#); Stairs et al. 2001), four of them are very high (>0.8; the exception being PSR J1740−3052, which still has a significant eccentricity of ∼0.6). This has been understood as being a result of a sizable velocity *kick* imparted to the neutron star at the time of the core-collapse supernova that produced it, due to asymmetries in the explosion. Such a kick will typically disrupt the binary, whereas for the known pulsar/massive-star binaries, it fortuitously did not. Other evidence for such kicks comes from the high space velocities of isolated radio pulsars relative to those of their progenitor population (O and B stars, e.g., Hobbs et al. [2005](#)), as well as from the observation of spin–orbit coupling in three of the pulsar/B-star binaries (Kaspi et al. [1996](#); Madsen et al. [2012](#); Shannon et al. [2014](#)), which implies the neutron star's orbit is presently misaligned with the progenitor star's orbit, a feat that can only have been accomplished via a supernova velocity kick directed out of that plane.

By contrast, the PSR J2108+4516 binary system is remarkable as its orbital eccentricity is relatively small at $e = 0.087$ (Table [2](#)). Low orbital eccentricities in some Be/X-ray binaries, the likely descendants of high-mass radio pulsar binaries, have been suggested to be due to the neutron star having formed in an electron-capture supernova (Pfahl et al. [2002](#); Knigge et al. [2011](#)), which is expected to be more symmetric—hence produce only small kicks—than an iron core-collapse supernova. Electron-capture supernovae are expected in O/Ne/Mg-core stars with masses in the range 8–10 $M_\odot$ (Nomoto & Hashimoto [1988](#); Woosley et al. [2002](#)). However, the present-day minimum companion mass in the PSR J2108+4516 system is 11.7 $M_\odot$, which suggests that the progenitor of the pulsar may have been more massive than 11.7 $M_\odot$, hence outside the expected mass range for an electron-capture explosion. As described in Section [5.4](#), it seems unlikely that significant mass transfer is occurring today. Such mass transfer would be expected if a previously highly eccentric orbit had been circularized. Hence, the present-day low eccentricity would have had to be a result of either a small magnitude kick in a conventional iron core-collapse or a larger kick having been imparted opposite to the progenitor orbital velocity, such that the two roughly canceled. The existence of the PSR J2108+4516 binary system is thus potential evidence that at least some of the low-eccentricity Be/X-ray binaries could also be the results of iron core-collapse supernovae.

On the other hand, mass transfer earlier on in the evolution of the binary could have resulted in mass added to the current companion, with its pretransfer mass having been significantly lower (see, e.g., Bhattacharya & van den Heuvel [1991](#)). In that case, the pulsar progenitor, just prior to collapse, may have been much less massive than the companion is today, and within the range for which an electron-capture collapse was possible. For example, as shown by Bhattacharya & van den Heuvel ([1991](#); see their Figure 25), for conservative mass transfer, a pulsar progenitor of initial mass 13 $M_\odot$ with a 6.5 $M_\odot$ companion can result in a system similar to that of PSR J2108+4516, with a 2.5 $M_\odot$ presupernova pulsar progenitor and a 17 $M_\odot$ companion. If future optical observations were to reveal that the present companion mass in the PSR J2108+4516 system is much greater than 17 $M_\odot$, an early mass transfer scenario might be less likely, arguing against the electron-capture possibility.

## 6. Conclusions

We have presented the CHIME/FRB discovery and 2.8 yr 2.8 yr CHIME/Pulsar timing of a new radio pulsar/massive-star binary, PSR J2108+4516, only the sixth such binary pulsar known. The pulsar, in a ∼269 day, 0.09-eccentricity orbit with a companion of minimum mass ∼11 $M_\odot$, undergoes episodic disappearances lasting weeks to months, as well as significant variations in DM (as large as ∼3.7(4) pc cm$^{-3}$) and scattering time (as large as ∼19(5) ms). These variations are likely due to intervening material from a circumstellar disk and/or very dense stellar wind associated with the companion star. Using observations with the VLA to localize the pulsar, we have identified the companion as EM* UHA 138, a $m_V \simeq 11$ mag OBe star located in the North America nebula of the Cygnus region, at a distance of 3 kpc (consistent with the measured average pulsar DM of 83.5 pc cm$^{-3}$). From the stellar magnitude, we infer a mass range of 17–23 $M_\odot$, although we note that the mass function from timing combined with the substantial eclipsing implies a mass lower than this range, closer to the minimum mass limit of ∼11 $M_\odot$. We suggest that the low binary eccentricity, remarkable among pulsar/B-star binaries, results from one of two possible pulsar formation scenarios: (1) a low-amplitude or fortuitously directed kick from a core-collapse supernova, or (2) a low-amplitude kick from an electron-capture supernova, preceded by a period of mass transfer from the pulsar progenitor to the companion.

PSR J2108+4516 promises to serve as another rare laboratory for the exploration of massive-star winds and/or circumstellar disks. Future exploratory observations include the following: optical spectroscopy to determine companion type and investigate whether it has a disk; X-ray and/or gamma-ray observations to study disk and wind interactions; very long baseline interferometry observations to constrain the pulsar orbit on-sky combined with Gaia astrometry to constrain the optical orbit, resulting in direct mass measurements; radio polarimetry to further characterize the clumpy circumstellar medium; and long-term radio timing, especially at higher frequencies, to study orbital dynamics and potential deviations from a Keplerian orbit due to spin–orbit coupling. Higher frequencies may be able to fill in the orbital phases of consistent nondetection, leading to more accurate determination of binary parameters.

We acknowledge that CHIME is located on the traditional, ancestral, and unceded territory of the Syilx/Okanagan people. We are grateful to the staff of the Dominion Radio Astrophysical Observatory, which is operated by the National Research Council of Canada. CHIME is funded by a grant from the Canada Foundation for Innovation (CFI) 2012 Leading Edge Fund (Project 31170) and by contributions from the provinces of British Columbia, Québec, and Ontario. The CHIME/FRB Project, which enabled development in common with the CHIME/Pulsar instrument, is funded by a grant from the CFI 2015 Innovation Fund (Project 33213) and by contributions from the provinces of British Columbia and Québec, and by the Dunlap Institute for Astronomy and Astrophysics at the University of Toronto. Additional support was provided by the Canadian Institute for Advanced Research (CIFAR), McGill University and the McGill Space Institute, the Trottier Family Foundation, and the University of British Columbia (UBC). The CHIME/Pulsar instrument hardware was funded by NSERC RTI-1 grant EQPEQ 458893-2014. Pulsar research at UBC is funded by an NSERC Discovery grant and by CIFAR. This research was enabled in part by support provided by WestGrid (www.westgrid.ca) and Compute Canada (www.computecanada.ca). The National Radio Astronomy Observatory is a facility of the National Science Foundation operated under cooperative agreement by Associated Universities, Inc. This research has made use of the SIMBAD and VizieR databases, operated at Centre de Données astronomiques de Strasbourg (CDS), Strasbourg, France. This work has made use of data from the European Space Agency (ESA) mission Gaia (https://www.cosmos.esa.int/gaia), processed by the Gaia Data Processing and Analysis Consortium (DPAC; https://www.cosmos.esa.int/web/gaia/dpac/consortium). Funding for the DPAC has been provided by national institutions, in particular the institutions participating in the Gaia Multilateral Agreement.

A.B.P. is a McGill Space Institute (MSI) Fellow and a Fonds de Recherche du Quebec—Nature et Technologies (FRQNT) postdoctoral fellow. B.C.A. is supported by an FRQNT Doctoral Research Award. C.L. was supported by the U.S. Department of Defense (DoD) through the National Defense Science and Engineering Graduate Fellowship (NDSEG) Program. F.A.D. is supported by the UBC four year fellowship. J.W.M. gratefully acknowledges support by the Natural Sciences and Engineering Research Council of Canada (NSERC), [funding reference #CITA 490888-16]. K.C. is supported by a UBC Four Year Fellowship (6456). K.W.M. holds the Adam J. Burgasser Chair in Astrophysics and is supported by an NSF grant (2008031). Pulsar research at UBC is supported by an NSERC Discovery grant and by the Canadian Institute for Advanced Research. S.P.T. is a CIFAR Azrieli Global Scholar in the Gravity and Extreme Universe Program. The National Radio Astronomy Observatory is a facility of the National Science Foundation operated under cooperative agreement by Associated Universities, Inc. S.M.R. is a CIFAR Fellow and is supported by the NSF Physics Frontiers Center awards 1430284 and 2020265. V.M.K. holds the Lorne Trottier Chair in Astrophysics & Cosmology, a Distinguished James McGill Professorship, and receives support from an NSERC Discovery grant (RGPIN 228738-13), from an R. Howard Webster Foundation Fellowship from CIFAR, and from the FRQNT CRAQ. Z.P. is a Dunlap Fellow.

We give special recognition to the memory of our colleague, Jing Luo, who enthusiastically contributed so much to this project, and who would have loved to have seen this result.

*Software*: VizieR (Ochsenbein et al. 2000), SIMBAD (Wenger et al. 2000), dspsr (van Straten & Bailes 2011), psrchive (Hotan et al. 2004; van Straten et al. 2012), tempo (Nice et al. 2015), CASA (McMullin et al. 2007), PulsePortraiture (Pennucci et al. 2014; Pennucci 2019), PsrSigSim (Shapiro-Albert et al. 2021).

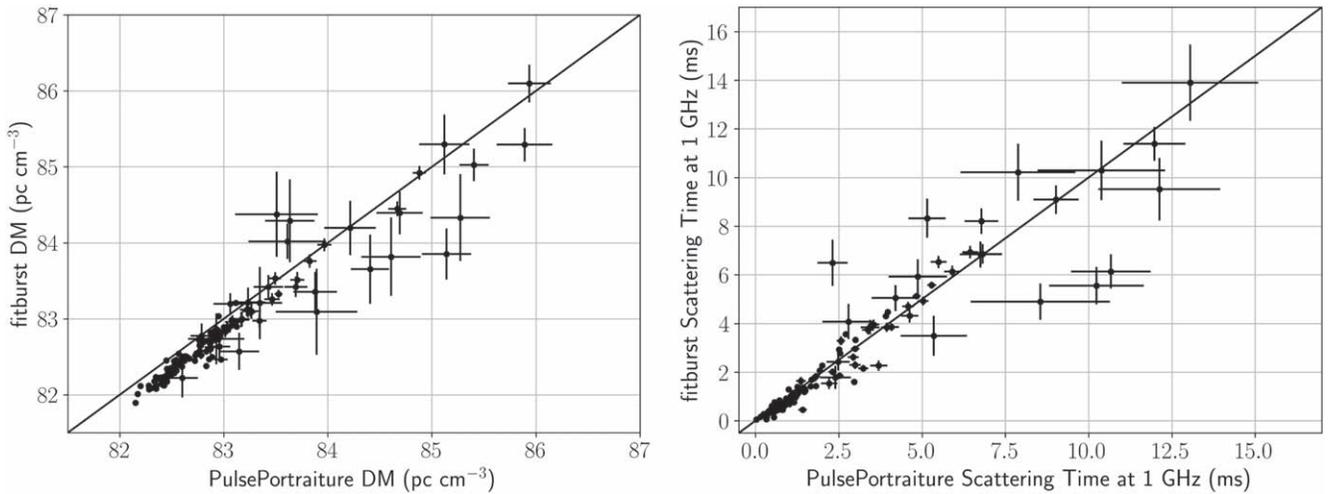

**Figure 10.** A comparison between `PulsePortraiture` and `fitburst` measurements for (left) DM and (right) scattering delay for a subset of 150 PSR J2108 +4516 observations. The thick black line indicates $y = x$. Error bars on the DM and scattering values represent $1\sigma$.

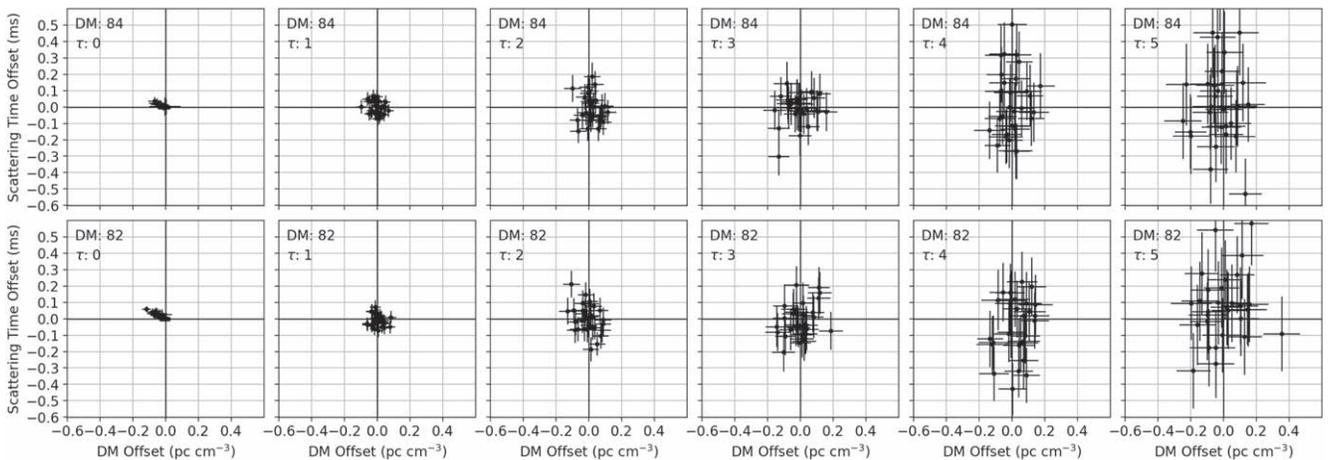

**Figure 11.** A summary of the offset of `PulsePortraiture` DM and scattering delay measurements from the intrinsic values simulated in pulse profile data using `PsrSigSim`. Each plot shows the results for the 30 noise realizations, with the offset in DM on the $x$-axis and the offset in scattering time on the $y$-axis. The intrinsic DM and scattering values are labeled in the text on each plot (by *DM* in parsecs per cubic centimeter and $\tau$ in milliseconds, respectively). Note that scattering times on these plots are referenced to 1 GHz.

## Appendix
## Pulse Portraiture Tests

In this Appendix, we briefly present three tests of `PulsePortraiture`'s ability to fit DM and scattering variations: (1) we compare `PulsePortraiture` DM–scattering measurements for PSR J2108+4516 to those obtained from the profile fitting software `fitburst` (Masui et al. 2015; CHIME/FRB Collaboration et al. 2021); (2) we use the `PsrSigSim` software package (Shapiro-Albert et al. 2021) to simulate PSR J2108 +4516-like CHIME/Pulsar data for a range of DM–scattering values, run `PulsePortraiture` on the simulations, and compare the recovered values to the intrinsic simulated values; (3) we complete another set of simulation comparisons, this time using `PulsePortraiture`'s `make_fake_pulsar` function to simulate data in the high-scattering regime.

### A1. Fitburst Comparison

For a subset of 150 observations, we ran the profile fitting software `fitburst` on the CHIME/Pulsar observation-integrated profiles to obtain DM and scattering measurements. `fitburst` directly models the 2D pulse profile using a least-squares fitting algorithm in the phase domain (rather than the Fourier domain, like `PulsePortraiture`). Similarly to `PulsePortraiture`, the temporal shape of the burst is modeled by a Gaussian intrinsic profile convolved with a one-side decaying exponential function to encapsulate any scattering in the pulse profile. In the frequency dimension, the spectral shape of the burst is fit by a continuous power-law function with a spectral index of $\alpha$ and an extra *spectral running* parameter $\beta$: $I(\nu) = (\nu/\nu_{\rm ref})\exp(\alpha + \beta \ln(\nu/\nu_{\rm ref}))$, for some reference frequency $\nu_{\rm ref}$. This flexible function allows the spectral profile to vacillate between a regular, broadband power law and a band-limited Gaussian, covering the wide variety of spectral shapes. More details can be found in CHIME/FRB Collaboration et al. (2021).

A comparison between the resulting `PulsePortraiture` and `fitburst` values is shown in Figure 10. Both the DM and scattering values roughly follow the $y = x$ relationship indicating agreement, with some residual scatter that is likely due to differences in the preprocessing and RFI removal methods. The ∼0.15 pc cm$^{-3}$ offset between the `PulsePortraiture` and `fitburst` DM values arises from differences in the modeling of the pulse profile as a function of frequency. This effect has been previously observed in comparisons between `PulsePortraiture` DMs and DMs determined from conventional narrow-band DMX fits (see Figure 6 of Pennucci et al. (2014) and accompanying discussion).

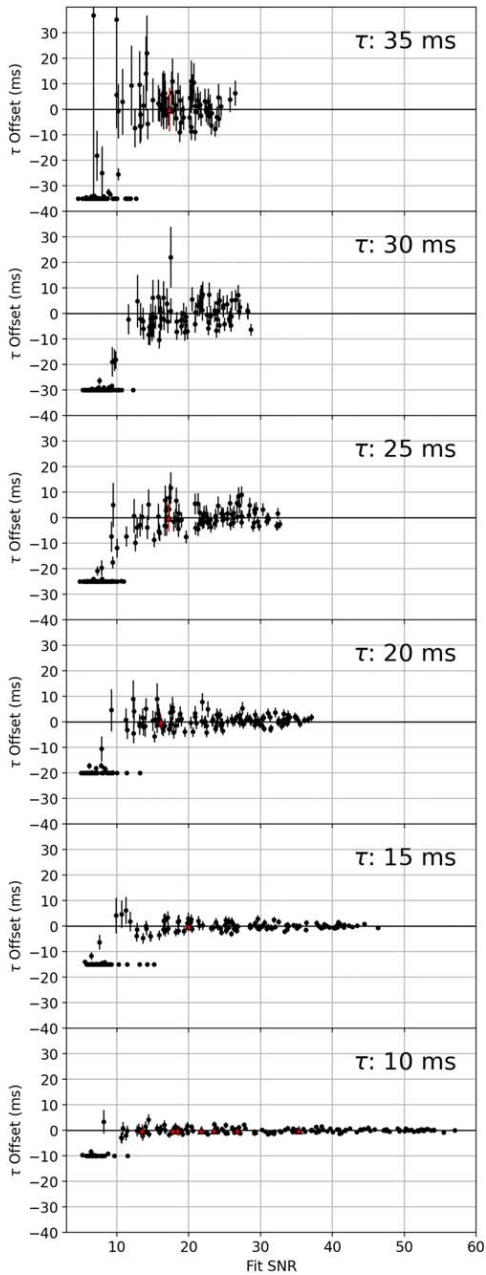

**Figure 12.** A summary of the offset of `PulsePortraiture` scattering delay measurements from the intrinsic values in the high-scattering regime. Profile data is simulated using the `PulsePortraiture` pplib module. Each plot shows 160 noise realizations at S/Ns ranging from the median to the lowest observed in our PSR J2108+4516 data set. The red triangles plotted on the $\tau$ offset zero line represent the fit S/Ns and scattering time uncertainties for the 12 real PSR J2108+4516 observations with scattering times >10 ms. Each of these points is plotted in the plot corresponding to the scattering delay that is closest to the measured value.

*A2. `PsrSigSim` Simulations*

We used the `PsrSigSim` software to simulate CHIME PSR J2108+4516 observations with different intrinsic scattering times and DMs covering the range of values expressed by the majority of PSR J2108+4516 profiles (DMs between 82 and 84 pc cm$^{-3}$ and scattering delays between 0 and 5 ms at 1 GHz). For each DM–scattering time pairing, we simulated 30 noise realizations assuming the median S/N of PSR J2108+4516 detections (determined on the pulse profile summed in time and frequency, using the `psrchive` `psrstat` utility). Then we ran `PulsePortraiture` on this simulated data using the same methods described in Section 3.3.

Figure 11 shows the offsets of the fit `PulsePortraiture` DMs and scattering times compared to the intrinsic simulated values. Notably, changing the DM has no discernible impact on the results. However, as the scattering increases, the spread in the results increases as well, but the offset in scattering time is generally <0.6 ms for scattering values <5 ms. Most of our real PSR J2108+4516 observations have a fit scattering time less than 2 ms (the third column from the left). In addition, when the scattering is low, `PulsePortraiture` interprets some of the DM as scattering. This is consistent with the covariance values that we extracted from the `PulsePortraiture` fits, which indicate that when the scattering is lower, DM and scattering become more covariant.

*A3. High-scattering Simulations*

We completed another round of simulations to more closely examine `PulsePortraiture`'s accuracy in fitting profiles with scattering values greater than 10 ms at 1 GHz. This regime is relevant for 12 of our PSR J2108+4516 observations. For such high scattering values, the scattering tail is long enough to wrap around in phase in the CHIME band, causing subsequent profiles to fold over each other in the fold-mode data. Since `PsrSigSim` does not simulate this effect, we instead switched to simulating data using the `make_fake_pulsar` function in the `PulsePortraiture` pplib module.

Figure 12 shows the offsets of the `PulsePortraiture`-fit scattering times compared to the intrinsic simulated values, as a function of the S/Ns of the fits.[45] For each intrinsic scattering value, we simulate 160 different profile S/Ns ranging from the median to the lowest observed in our PSR J2108+4516 data set (again, determined using `psrchive`'s `psrstat`, as in Appendix A2). Based on these fits, it appears that `PulsePortraiture` largely recovers the intrinsic scattering value within the uncertainty. However, there are a few instances when `PulsePortraiture` can overestimate or underestimate the scattering time by up to ∼10 ms at scattering times ⩾20 ms. Below a fit S/N of 10, the results become more unreliable. Notably, there are instances where the fit latches onto a noise fluctuation, resulting in a scattering measurement of 0. However, none of our 12 PSR J2108+4516 observations with scattering times >10 ms fall within this fit S/N range.

---

[45] `PulsePortraiture` outputs an snr parameter that indicates the significance of each fit. This parameter is given by the following:

$$\mathrm{snr} = \sqrt{\sum_n \frac{C_{dp,n}}{(S_{p,n})^{1/2}}},$$

where $n$ is the number of frequency channels, and $C_{dp,n}$ and $S_{p,n}$ are defined in Equations (10)b and (10)c in Pennucci et al. (2014), respectively.

## ORCID iDs

Bridget C. Andersen https://orcid.org/0000-0001-5908-3152
Emmanuel Fonseca https://orcid.org/0000-0001-8384-5049
J. W. McKee https://orcid.org/0000-0002-2885-8485
B. W. Meyers https://orcid.org/0000-0001-8845-1225
Jing Luo https://orcid.org/0000-0001-5373-5914
C. M. Tan https://orcid.org/0000-0001-7509-0117
I. H. Stairs https://orcid.org/0000-0001-9784-8670
Victoria M. Kaspi https://orcid.org/0000-0001-9345-0307
M. H. van Kerkwijk https://orcid.org/0000-0002-5830-8505
Mohit Bhardwaj https://orcid.org/0000-0002-3615-3514
P. J. Boyle https://orcid.org/0000-0001-8537-9299
Kathryn Crowter https://orcid.org/0000-0002-1529-5169
Paul B. Demorest https://orcid.org/0000-0002-6664-965X
Fengqiu A. Dong https://orcid.org/0000-0003-4098-5222
Deborah C. Good https://orcid.org/0000-0003-1884-348X
Jane F. Kaczmarek https://orcid.org/0000-0003-4810-7803
Calvin Leung https://orcid.org/0000-0002-4209-7408
Kiyoshi W. Masui https://orcid.org/0000-0002-4279-6946
Arun Naidu https://orcid.org/0000-0002-9225-9428
Cherry Ng https://orcid.org/0000-0002-3616-5160
Chitrang Patel https://orcid.org/0000-0003-3367-1073
Aaron B. Pearlman https://orcid.org/0000-0002-8912-0732
Ziggy Pleunis https://orcid.org/0000-0002-4795-697X
Masoud Rafiei-Ravandi https://orcid.org/0000-0001-7694-6650
Mubdi Rahman https://orcid.org/0000-0003-1842-6096
Scott M. Ransom https://orcid.org/0000-0001-5799-9714
Kendrick M. Smith https://orcid.org/0000-0002-2088-3125
Shriharsh P. Tendulkar https://orcid.org/0000-0003-2548-2926